\documentclass[ALICE,manyauthors]{cernphprep}
\newcommand{\pt}{p_\mathrm{T}} % \pt, transverse momentum
\newcommand{\ptmu}{p_{\mathrm{T},\mu}} % \pt, transverse momentum
 
\newcommand{\npart}{\left\langle N_\mathrm{part}\right\rangle}
\usepackage[comma,square,numbers,sort&compress]{natbib}
\usepackage{hyperref}
\usepackage{epstopdf}
\usepackage{lineno}
\usepackage{array}
\usepackage{longtable}
\usepackage{multirow}
\usepackage{ctable}
\usepackage{booktabs}
\usepackage{color}
\usepackage{textcomp}
\begin{document}%

%%%%%%%%%%%%%%%  Title page %%%%%%%%%%%%%%%%%%%%%%%%%
\begin{titlepage}
\PHyear{2018}
\PHnumber{082}      % required, will be obtained from PH
\PHdate{21 April}  % required, will be obtained from PH
%

%%% Put your own title + short title here:
\title{$\phi$ meson production at forward rapidity\\
       in Pb--Pb collisions at $\sqrt{s_\mathrm{NN}}=2.76$~TeV}
\ShortTitle{$\phi$ meson production at forward rapidity in Pb--Pb collisions at 
       $\sqrt{s_\mathrm{NN}}=2.76$~TeV}   % appears on right page headers

%%% Do not change the next lines
\Collaboration{ALICE Collaboration\thanks{See Appendix~\ref{app:collab} for the list of collaboration members}}
\ShortAuthor{ALICE Collaboration} % appears on left page headers, do not change

\begin{abstract}
$\phi$ meson measurements provide insight into strangeness production, which is one of 
the key observables for the hot medium formed in high-energy heavy-ion 
collisions. ALICE measured $\phi$ production through its decay 
in muon pairs in Pb--Pb collisions at $\sqrt{s_\mathrm{NN}}$ = 2.76 TeV 
in the intermediate transverse momentum range $2 < \pt < 5$~GeV/\textit{c} 
and in the rapidity interval $2.5<y<4$. 
The $\phi$ yield was measured as a function of the transverse momentum and collision centrality. 
The nuclear modification factor was obtained as a
function of the average number of participating nucleons.  
Results were compared with the ones obtained via the kaon decay channel in the same 
$\pt$ range at midrapidity. The values of the nuclear modification factor in the two rapidity 
regions are in agreement within uncertainties.

\end{abstract}
\end{titlepage}
\setcounter{page}{2}

\section{Introduction}
\label{sect:Introduction}

%The theory of Quantum Chromodynamics (QCD) predicts the occurrence of a phase transition from the hadronic matter to a plasma of deconfined quarks and gluons (Quark-GluonPlasma, QGP) at extreme conditions of temperature and energy density. 
At small values of the baryochemical potential and at extreme high temperatures, Quantum Chromodynamics (QCD) predicts chiral and deconfinement crossover transitions from hadronic matter to a state of strongly interacting medium, where dominant degrees of freedom are gluons and light quarks (Quark-Gluon Plasma, QGP).
Ultrarelativistic heavy-ion collisions provide the tools to study this phase of 
matter in the laboratory. Strangeness production is a key tool to understand the properties 
of the medium formed in these collisions. Indeed, an enhanced production of strange particles with respect to elementary hadronic collisions was early proposed as one of the signatures of the QGP~\cite{Koch:1986ud}.  
This enhancement is currently interpreted as resulting from the restoration of the chemical equilibrium between u, d and s quarks in sufficiently central heavy-ion collisions, with respect to ee and pp interactions, where strangeness production is expected to be canonically suppressed~\cite{Tounsi:2001ck}.

The $\phi$ meson, due to its $s \bar s$ valence quark content, provides insight 
into strangeness production. Since its cross section for interactions with 
non-strange hadrons can be assumed to be small, the $\phi$ meson should be less affected by 
hadronic rescattering during the expanding hadronic phase, which follows the QGP phase. %after hadronization in the system evolution, 
For this reason, the $\phi$ meson better reflects the early evolution of the 
system~\cite{PhysRevLett.54.1122}. 
Because of the long lifetime of the $\phi$ meson, the rescattering effects that should 
affect the hadronic decay channels are negligible~\cite{PhysRevLett.96.152301,ARNALDI2011325,Abelev:2014uua,PhysRevC.93.014911}, 
making thus possible a direct comparison between the hadronic and dileptonic decay channels.

Moreover, the $\phi$ meson may be sensitive to chiral 
symmetry restoration~\cite{LISSAUER199115,Aoki200646,PhysRevD.74.054507}, that could be observed by measuring a mass shift 
of a few MeV/$c^2$ or a broadening of the spectral function of the hadronic resonances up to several times their 
PDG value~\cite{Klingl1998254,Johnson2001,Rapp:2009yu,Eletsky:2001bb}. %~\cite{Klingl1998254,Johnson2001,Brown:2001nh, Rapp:2009yu, Eletsky:2001bb,Gubler:2016itj}. 
However, no experimental evidence of such a broadening or mass shift has been  
observed so far for the $\phi$ meson in high-energy heavy-ion collisions neither in the hadronic nor in the 
dilepton decay channel~\cite{Alt:2008iv,Banicz:2009aa,PhysRevC.79.064903,PhysRevC.92.024912,PhysRevC.72.014903,Abelev:2014uua}. 
%, although KEK-PS E325 experiment observed a mass modification of the $\phi$ meson at normal %nuclear density in p-Cu collisions at 12 GeV, in the dilepton channel $e^+ e^-$~\cite{PhysRevLett.98.042501,PhysRevLett.98.152302}.
 
The measurement of hadrons in different $\pt$ ranges provides important information on the relative 
contribution of different possible hadronization mechanisms. Soft processes dominate the 
low transverse momentum region ($\pt \lesssim 2$~GeV/$c$), where the system evolution can 
be described on the basis of hydrodynamical models and particle yields follow
the expectations of thermal models~\cite{Cleymans:2006xj, Rafelski:2004dp, 
Petran:2013lja, Andronic:2008gu, Becattini:2010sk, Bozek:2012qs, Werner:2012xh, BRAUNMUNZINGER199543, BRAUNMUNZINGER19961}. 
%\textbf{In sufficiently central heavy ion collisions the yield of strange hadrons appears to be in chemical equilibrium, while it is suppressed in $e^+e^-$ and pp collisions??????????????}
On the other side, for high transverse momenta ($\pt \gtrsim 5$~GeV/$c$), hard parton-parton scattering processes and subsequent fragmentation become the dominant production mechanisms. In the presence of a deconfined medium, additionally, parton energy loss via elastic collisions and gluon bremsstrahlung~\cite{d'Enterria:2009am} modifies the spectral distributions, leading to a suppression of hadron production in central heavy-ion collisions 
with respect to the one measured in peripheral heavy-ion or in pp collisions, scaled by the number of binary collisions. 

At intermediate transverse momenta ($2 < \pt < 5$~GeV/\textit{c}), measurements at 
RHIC showed an enhancement above unity of the ratio between the baryon and meson yields, 
the so-called ``baryon anomaly". This has been attributed to the recombination of quarks~\cite{PhysRevC.67.034902, PhysRevLett.90.202303, PhysRevLett.90.202302, Adcox:2001mf, Adler:2003kg, Adams:2003am}.
However, measurements at the LHC~\cite{Abelev:2014laa}
showed that the proton-to-pion ratio from low to intermediate $\pt$ could be described 
by hydrodynamical models~\cite{Bozek:2012qs, Werner:2012xh}.
The $\phi$, being a meson and having a mass close to that of the proton, 
is an ideal probe to disentangle whether this effect is more
related to the particle mass or to its valence quark content, since recombination scales with the number of quarks, 
while hydrodynamical models depend on the particle mass.
%At the LHC, hydrodynamical models seem to describe the data even in the intermediate region, and the radial flow 
%is seen to play an important role in the $p_T$ distributions. A test of the models at forward rapidity 
%would complement the results already obtained at midrapidity and give an insight 
%to the effect of radial flow in a wide rapidity range. 

%At the LHC energies, hydrodynamical models seem to work even in the intermediate region, and the radial flow is seen to play an important role in the pT distribution of the particles. It should be interesting to check the effect of the radial flow in central and forward region and if there are strong different betwwen the production process at forward and midrapidity.
Recent measurements at the LHC~\cite{Abelev:2014uua} 
showed that the $p/\phi$ ratio at midrapidity does not show a significant dependence 
on $\pt$, while the $p/\pi$ and $\phi/\pi$ ratios 
show similar increases as a function of the transverse momentum, 
indicating that particle radial flow and therefore the particle masses
mainly determine the $\pt$ distributions of these particles. 
Hence, it is interesting to test whether there is a dependence of radial flow on rapidity 
%, which is expected to be more important at midrapidity than at forward rapidity~\cite{ARSENE201022}, 
and to compare the results at forward and midrapidity within the same experiment.
A comparison with hydrodynamical models at forward rapidity 
would complement the results already obtained at midrapidity, where they have shown to describe the 
data even in the intermediate $\pt$ region.
% and give an insight to the effect of radial flow in a wide rapidity range. 

This article presents a measurement of $\phi$ production in Pb--Pb collisions at 
$\sqrt{s_\mathrm{NN}}=2.76$~TeV at forward rapidity with the ALICE muon spectrometer at the LHC. The $\phi$ meson was reconstructed in the rapidity range $2.5 < y < 4$ for intermediate transverse momenta in the range $2 < \pt < 5$~GeV/$c$ via its decay in muon pairs. 

%To address the search for in-medium modifications of the $\phi$, it is important to measure them via the leptonic decay channel in addition to the dominant hadronic channel, because dileptons are not influenced by final state interactions, while kaons are unlikely to escape the fireball without reacting further, thus destroying any useful information of the spectral function of the $\phi$.

% The evolution of the $\phi$ yield with centrality and transverse momentum 
% is discussed and compared with the measurement at midrapidity in 
% the kaon decay channel~\cite{Abelev:2014uua}. Finally, nuclear modification factors are determined. 

The evolution of the $\phi$ yield with centrality and transverse momentum is discussed and compared with the measurement at midrapidity in the kaon decay channel~\cite{Abelev:2014uua}.  
Finally, the nuclear modification factors are determined. 

\section{Experimental apparatus}
\label{sect:ExperimentalApparatus}

The ALICE detector is described in detail in~\cite{Aamodt:2008zz}. The detectors relevant for this analysis are the forward muon spectrometer, the V0 detector, the Silicon Pixel Detector (SPD) of the Inner Tracking System (ITS) and the Zero Degree Calorimeters (ZDC). 

The muon spectrometer covers the pseudorapidity region $-4~ < \eta <-2.5$~\footnote{In the ALICE reference frame the muon
spectrometer covers negative $\eta$. However, we use positive values when referring to $y$.}; its elements are a front hadron absorber, followed by a set of tracking stations, a dipole magnet, an iron wall acting as muon filter and a trigger system. The front hadron absorber is made of carbon, concrete and steel and is placed at a distance of 0.9~m from the nominal interaction point (IP). 
Its total length of material corresponds to ten hadronic interaction lengths.
The 5~m long dipole magnet provides a magnetic field of up to 0.7~T in 
the vertical direction, which results in a field integral of 3~T$\cdot$m. A set 
of five tracking stations, each one composed of two cathode pad chambers, provides 
the muon tracking. The stations are located between 5.2 and 14.4~m from the IP, the 
first two upstream of the dipole magnet, the third in the middle of the dipole 
magnet gap and the last two downstream of it. 
%The intrinsic spatial resolution of the tracking chambers is $\sim$100 $\mu$m in the bending direction. 
A 1.2~m thick iron wall, corresponding to 7.2 hadronic interaction lengths, is 
placed between the tracking and trigger systems and absorbs the residual secondary 
hadrons emerging from the front absorber. The front absorber together with the 
muon filter stops muons with momenta lower than $\sim$4~GeV/\textit{c}. The tracking 
apparatus is completed by a muon triggering system (MTR) consisting of two detector 
stations, placed at 16.1 and 17.1~m from the IP. Each station is composed of two planes of resistive plate chambers.

The V0 detector is composed of two arrays of 32 scintillator sectors placed at 3.4~m and $-0.9$~m from 
the IP and covering the pseudorapidity regions 2.8~$< \eta <$~5.1 (V0A) and 
$-3.7 < \eta < -1.7$ (V0C), respectively. It is used to reject the background from beam-gas interactions and estimate the collision centrality and event plane.
The SPD, used for the determination of the primary vertex position, consists of two cylindrical layers of silicon pixel detectors, positioned at a radius of 3.9 and 7.6 cm from the beam axis. The pseudorapidity range covered by the inner and the outer layers is $|\eta| <$~2.0 and $|\eta| <$~1.4, respectively. The ZDC are located at $\sim$114~m from the IP and cover the pseudorapidity region $|\eta| > 8.7$. 
In this analysis they are used to reject electromagnetic interactions of lead ion beams. %, by applying a cut on the minimum energy deposited by spectator neutrons in the ZDC. 

\section{Data analysis}
\label{sect:DataAnalysis}

The analysis presented in this paper is based on the data sample collected by 
ALICE in 2011 during the Pb--Pb run at $\sqrt{s_\mathrm{NN}}= 2.76$~TeV.

The minimum bias (MB) trigger is defined as the coincidence of a signal in V0A
and V0C, synchronized with the passage of two colliding lead bunches. 
Data were collected with a dimuon unlike-sign trigger ($\mu\mu$MB), which is defined as the 
coincidence of a MB trigger and at least a pair of opposite-sign 
(OS) tracks selected by the MTR system, each with a transverse momentum above the 
threshold\footnote{The trigger threshold is 
not at a sharp value, but defined here as the $\pt$ 
value for which the trigger probability is 50\%.},  $\ptmu \gtrsim 1$~GeV/\textit{c}.

%Miminum bias events are used to normalize the total number of events.
The background events coming from beam interactions with the residual gas were reduced offline using the timing information on signals from the V0 and from the ZDC~\cite{performance_report}. %physics selection

The number of OS dimuon triggers collected is 1.7~$\cdot$~10$^7$, 
corresponding to an integrated luminosity of
$L_\mathrm{int}=68.8\pm 0.9 \mathrm{(stat)}^{+6.0}_{-5.1}\mathrm{(syst)}~\mu$b$^{-1}$~\cite{Abelev:2013ila}. 

The centrality determination is performed by fitting a distribution obtained with the 
Glauber model approach to the V0 amplitude distribution~\cite{Abelev:2013qoq}. 
%The centrality determination is based on a fit of simulated distributions, obtained with the 
%Glauber model approach, to the distribution of the summed signal amplitudes as 
%measured with the two V0 detectors~\cite{Abelev:2013qoq}. 
In the centrality range 0--90\% the efficiency of the MB trigger is 100\%
and the contamination from electromagnetic processes is negligible.
Events cor\-re\-spon\-ding to the 90\% most central collisions 
were thus selected. The centrality classes considered in this analysis were 0--20\%, 20--40\%, 40--60\% and 60--90\%. 

The Glauber model fit to the V0 signal distribution also allows to extract variables related to 
the collision geometry, such as the average number of participating nucleons $\npart$ and the 
nuclear overlap function $\langle T_\mathrm{AA} \rangle$, as reported in Table~\ref{tab:npart}.
% for the four centrality classes considered in this analysis. 

\begin{table}[h!]
\begin{center}
     \begin{tabular}{|c|c|c|}
      \hline\hline
       Centrality bin & $\npart$ & $\langle T_\mathrm{AA} \rangle$ (mb$^{-1})$ \\
       \hline%\hline
       0--20\% & 308.10 $\pm$ 3.70 & 18.91 $\pm$ 0.61 \\
       %\hline
       20--40\% & 157.20 $\pm$ 3.10 & 6.85 $\pm$ 0.23 \\
       %\hline
       40--60\% & 68.56 $\pm$ 2.00 & 2.00 $\pm$ 0.10 \\
       %\hline
       60--90\% & 17.55 $\pm$ 0.72 & 0.31 $\pm$ 0.03 \\
       \hline
       0--90\% & 124.40 $\pm$ 2.20 & 6.27 $\pm$ 0.21 \\
       \hline\hline
     \end{tabular}
     \caption{Average number of participating nucleons $\npart$ and nuclear 
   overlap function $\langle T_\mathrm{AA} \rangle$ for each centrality class~\cite{Abelev:2013qoq}.}   
    \label{tab:npart}
 \end{center}
\end{table}

Muon tracks were selected requiring a single muon $\ptmu >$~0.85 GeV/\textit{c}, %consequence of the  
to reject muons with a transverse momentum much below the 
hardware $\ptmu$ threshold imposed by the trigger system.
The selection of the muon pseudorapidity $-4 < \eta_{\mu} < -2.5$ was applied in 
order to remove the tracks close to the acceptance borders. Tracks crossing the 
part of the front absorber with the highest material density were rejected by 
restricting the transverse radial coordinate of the track at the end of the absorber to the range 
$17.6 < R_\mathrm{abs} < 89.5$~cm. Each track reconstructed in the tracking chambers was required to 
match a track reconstructed in the trigger chambers.
%The selection on the track $\chi^2_{\mu} >$ 5 was applied too.

Dimuons were selected requiring that their rapidity was in the range 
2.5~$< y <$~4. The trigger threshold on the single muon transverse momentum strongly 
reduces the detection efficiency for low mass, low $\pt$ dimuons. Therefore, the 
analysis was limited to dimuon transverse momenta in the range 
$2<p_{\mathrm{T}}<5$~GeV/\textit{c}, where the upper limit is only set by the currently 
available statistics.

%In the mass region 0 $< M <$ 10 GeV/\textit{c$^2$}, the number of OS muon pairs 
%satisfying the selections was 1.99~$\cdot$~10$^6$.

The opposite-sign dimuon invariant mass spectrum consists of correlated and uncorrelated pairs. 
The latter come mostly from the decay of pions and kaons and constitute the combinatorial 
background, which was evaluated via an event mixing technique, described in detail in~\cite{ALICE:2011ad}.
Events were assigned to classes of similar vertex position, event plane orientation
and centrality. %Vertex and event plane are both measured by V0 detector. 
Pairs were then formed with muons coming from different events belonging to the 
same classes. In this way, the resulting invariant mass spectrum consists of muon 
pairs which are uncorrelated by construction. 
The mixed events mass spectra were normalized to $2R\sqrt{N_{++} N_{--}}$, where $N_{++}$ ($N_{--}$) 
is the number of like-sign positive (negative) pairs integrated in the full mass range.
% The $R$ factor is defined as $A_{+-}/\sqrt{A_{++}A_{--}}$, where $A_{+-}$ ($A_{++}, A_{--}$) 
% is the acceptance for a $+-$ ($++$, $--$) pair and takes into account the possible correlations introduced by the detector. 
The $R$ factor takes into account the differences between the acceptances for like-sign and opposite-sign muon pairs and 
was estimated as $R = N^\mathrm{mixed}_{+-} / (2 \sqrt{N^\mathrm{mixed}_{++} N^\mathrm{mixed}_{--}})$, 
where $N^\mathrm{mixed}_{\pm\pm}$ is the number of mixed pairs for a given charge combination.

The quality of the combinatorial background determination
was checked through a Monte Carlo (MC) simulation in which uncorrelated 
muon pairs were generated. The muon transverse momentum and rapidity distributions were
parametrized to reproduce those from the experimental data. 
The detector response for these pairs was obtained with a
simulation that uses GEANT3~\cite{Brun:1119728}. 
The simulation results were then subjected to the same reconstruction 
and selection chain as the real data. In this way, all the possible correlations 
introduced by the detector were properly taken into account. 
The event mixing technique was then applied to the simulated pairs. The resulting 
opposite-sign mass spectrum was compared to the corresponding one 
obtained from the muon pairs in the same event.  Differences within 
2\% in the two distributions were observed. The limited precision in the combinatorial background subtraction 
was taken into account in the evaluation of the systematic uncertainty, as described below. 
%As a cross-check, the mass spectra of the measured like-sign pairs were compared with the
%ones obtained with the event mixing technique. It is assumed that like-sign pairs are 
%uncorrelated. The two distributions agree within $2\%$ in each centrality class and $\pt$ bin, for $m>0.6$~GeV/$c^2$.

\begin{figure}[h!]
	\centering
		\includegraphics[width=0.49\textwidth]{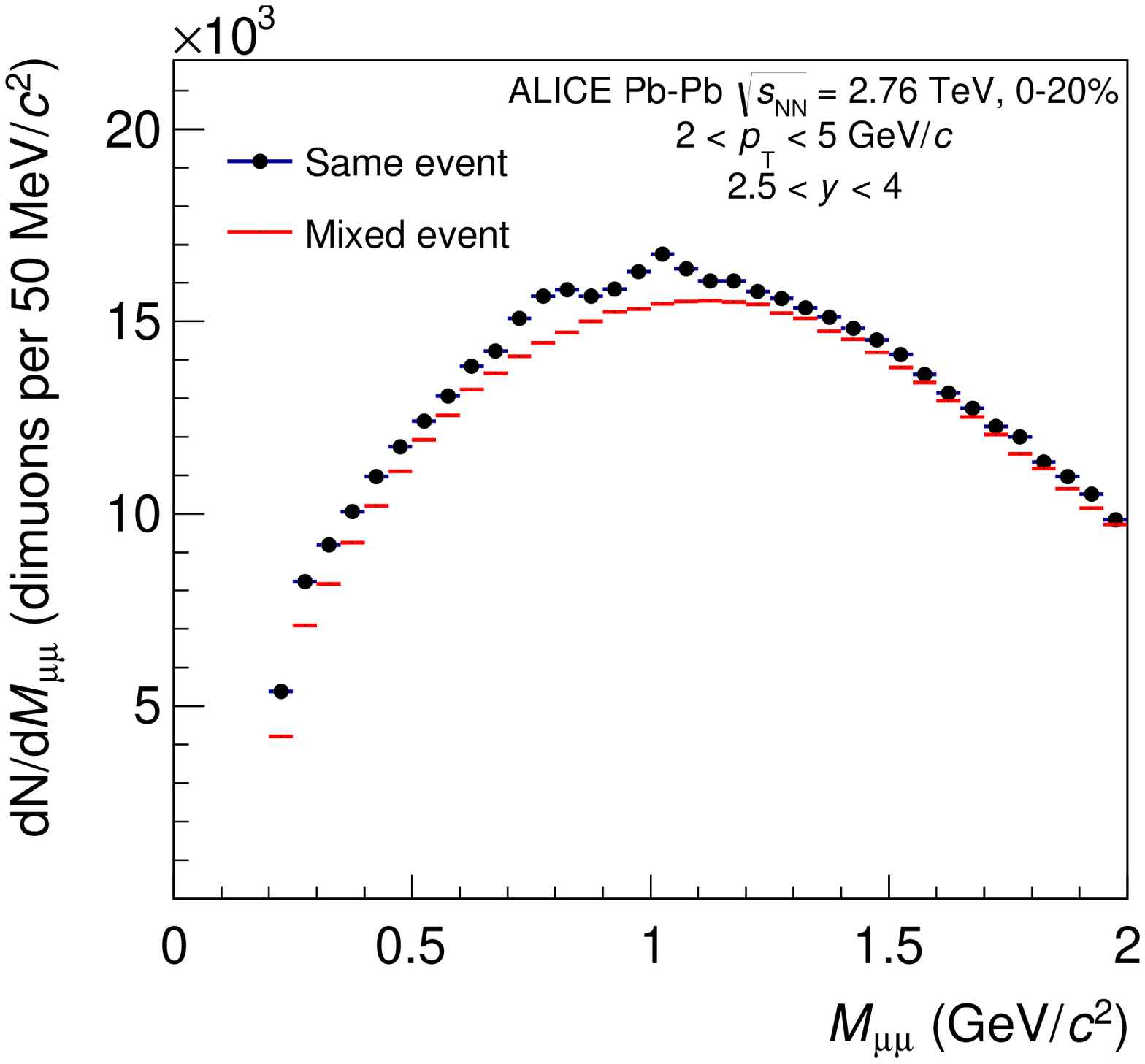}
		\includegraphics[width=0.49\textwidth]{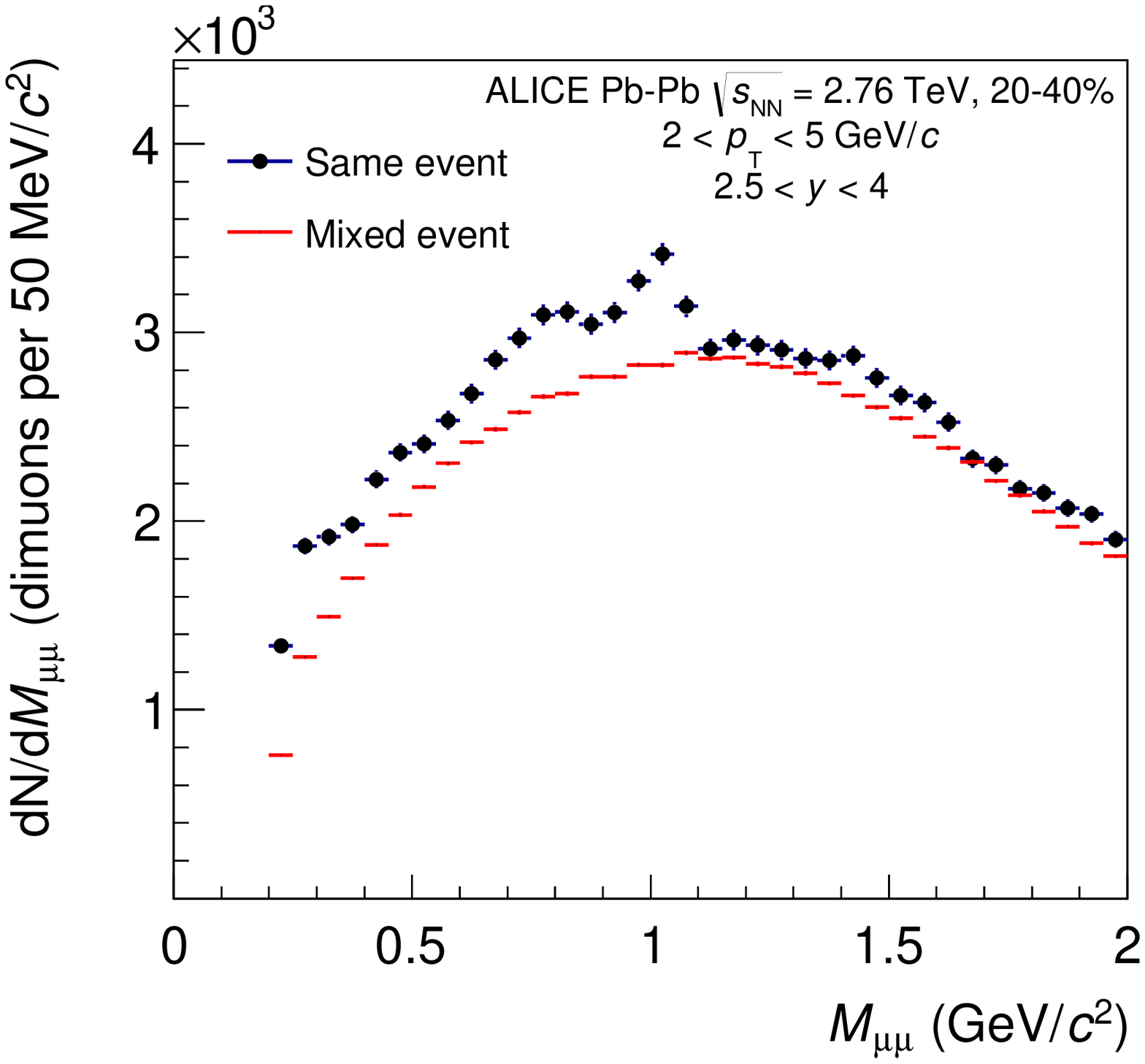}
                \includegraphics[width=0.49\textwidth]{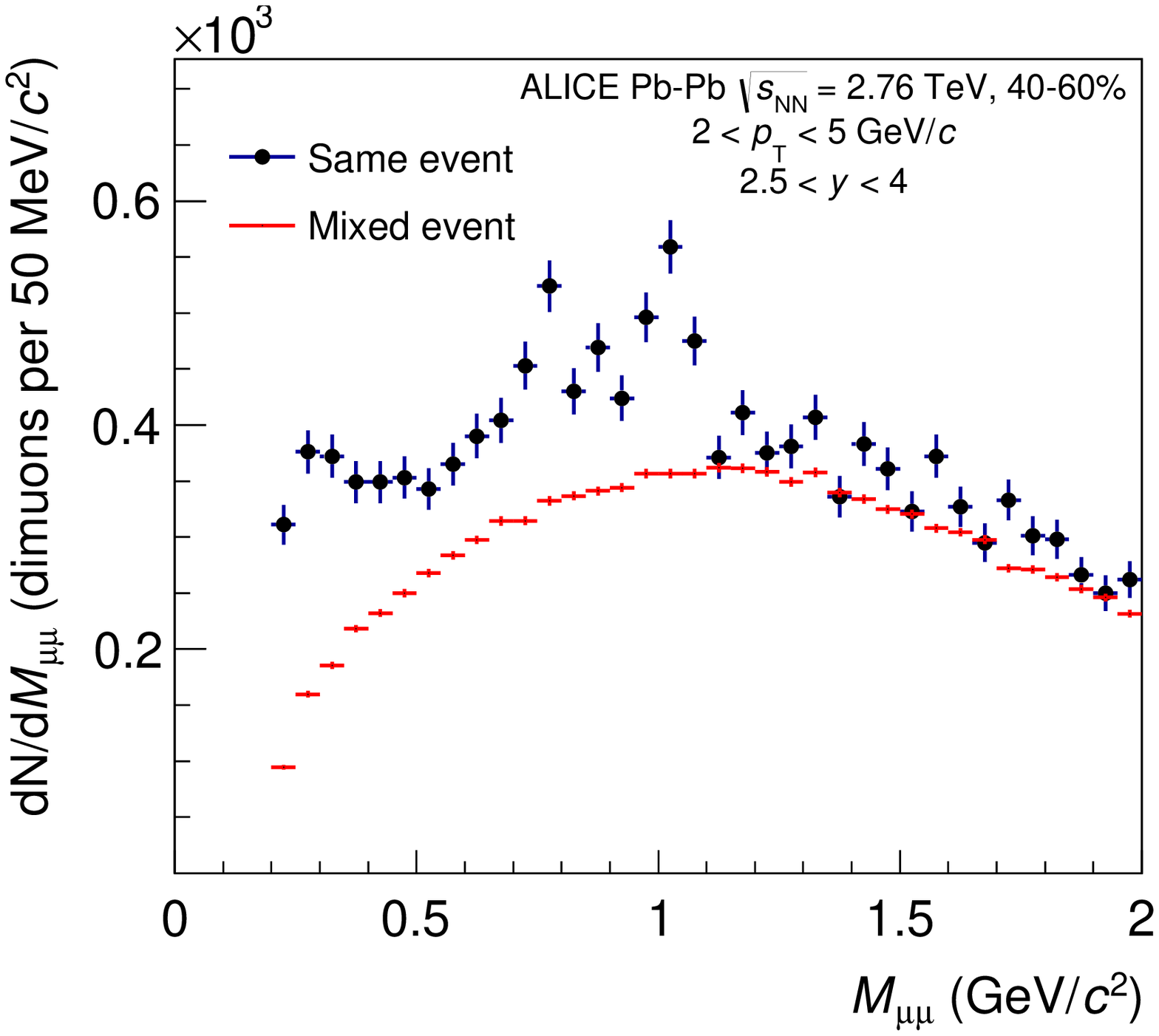}
                \includegraphics[width=0.49\textwidth]{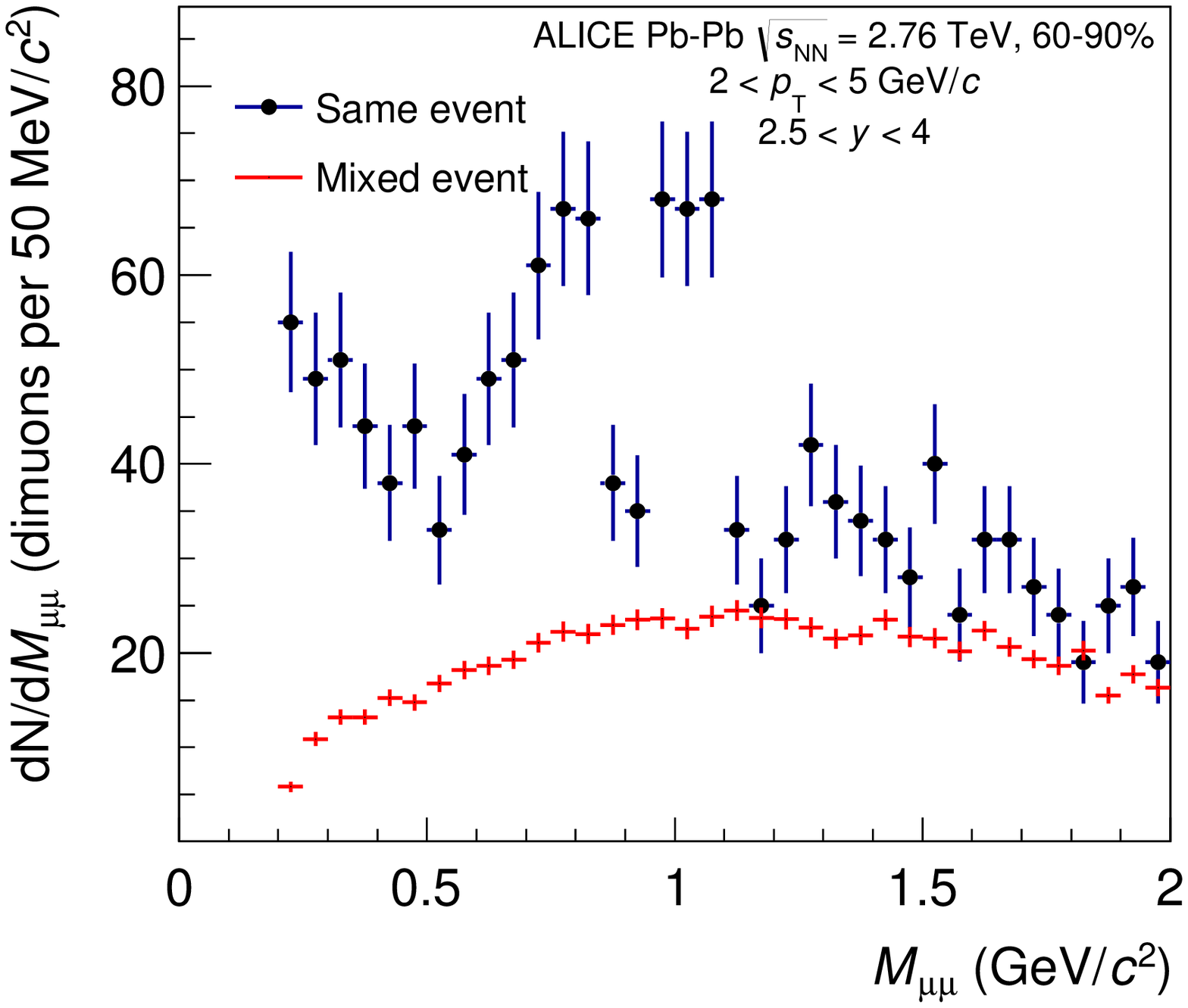}
	\caption{Invariant mass spectra for opposite-sign muon pairs in different centrality classes, in the range $2 < \pt < 5$ GeV/\textit{c}. The combinatorial background, evaluated from opposite-sign pairs in mixed events, is also shown.}
	\label{fig:os}
\end{figure}

Figure~\ref{fig:os} shows the invariant mass spectra for opposite-sign muon pairs in different centrality classes, before the combinatorial background subtraction, in the range $2 < \pt < 5$ GeV/\textit{c}. The combinatorial background, evaluated from opposite-sign pairs in mixed events, is also shown. 

\begin{figure}[h!]
	\centering
	\includegraphics[width=0.65\textwidth]{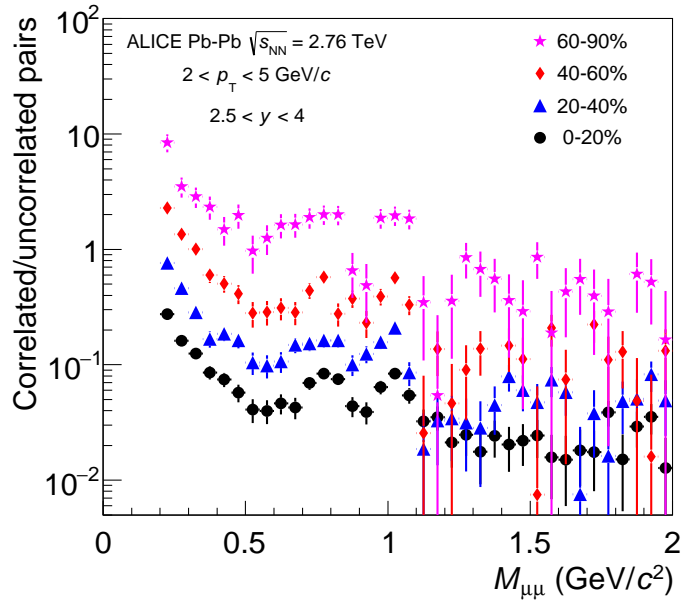}
	    \caption{Ratio between the mass spectra of correlated and uncorrelated pairs 
	    for different centralities, in the range 2~$< p_\mathrm{T} <$~5~GeV/\textit{c} 
	    in Pb--Pb collisions at $\sqrt{s_\mathrm{NN}}$ = 2.76~TeV.}
	\label{fig:sigbkg}
\end{figure}

The ratio between the invariant mass spectra of correlated and uncorrelated pairs for the different centralities is shown in Fig.~\ref{fig:sigbkg}: for $0.95 < M_{\mu\mu} < 1.1$~GeV/$c^2$ this ratio
increases from $\sim$0.07 in central collisions to $\sim$2 in peripheral collisions. 

\begin{figure}[h!]
	\centering
	\includegraphics[width=0.47\textwidth]{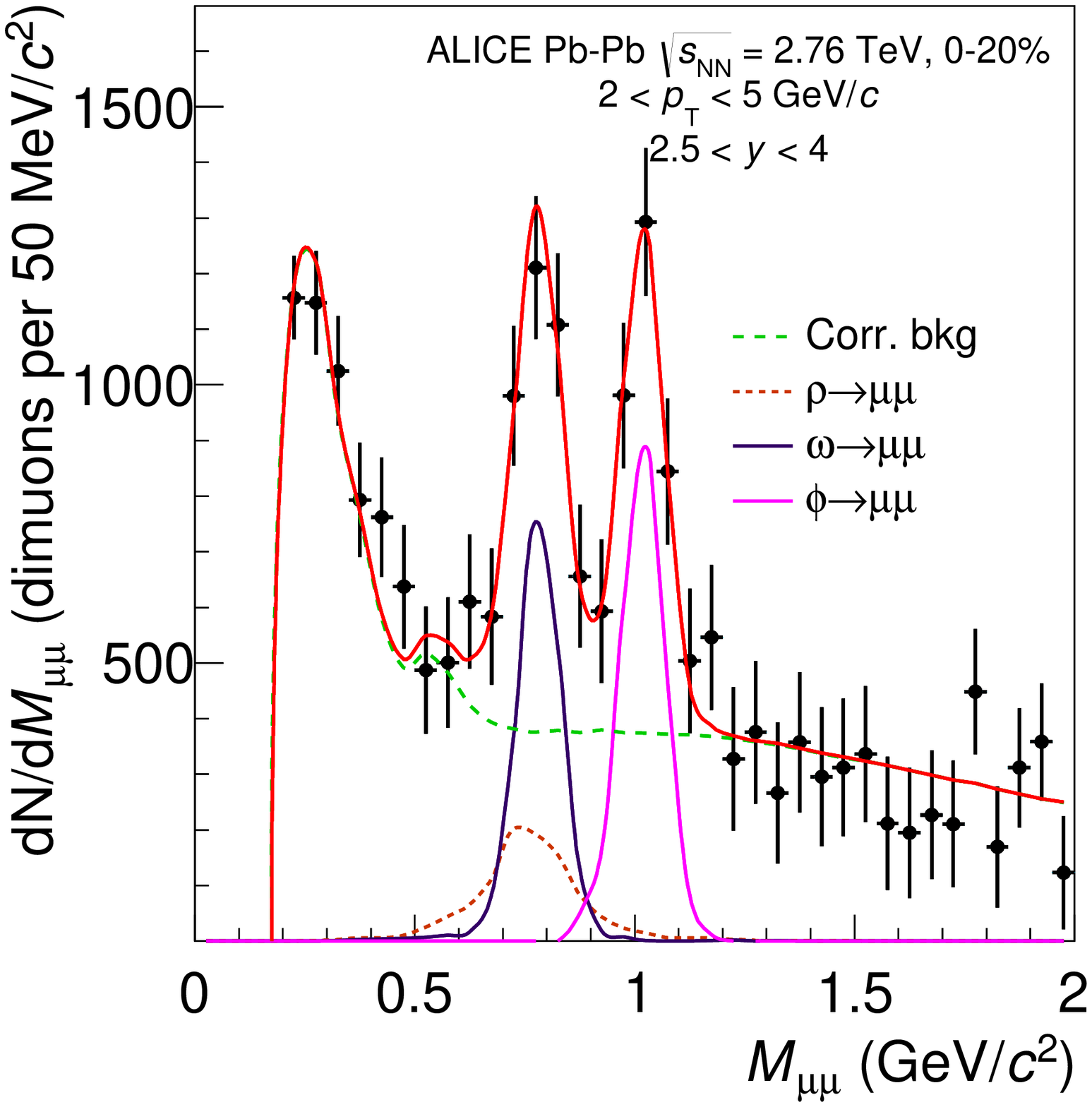}
	\includegraphics[width=0.47\textwidth]{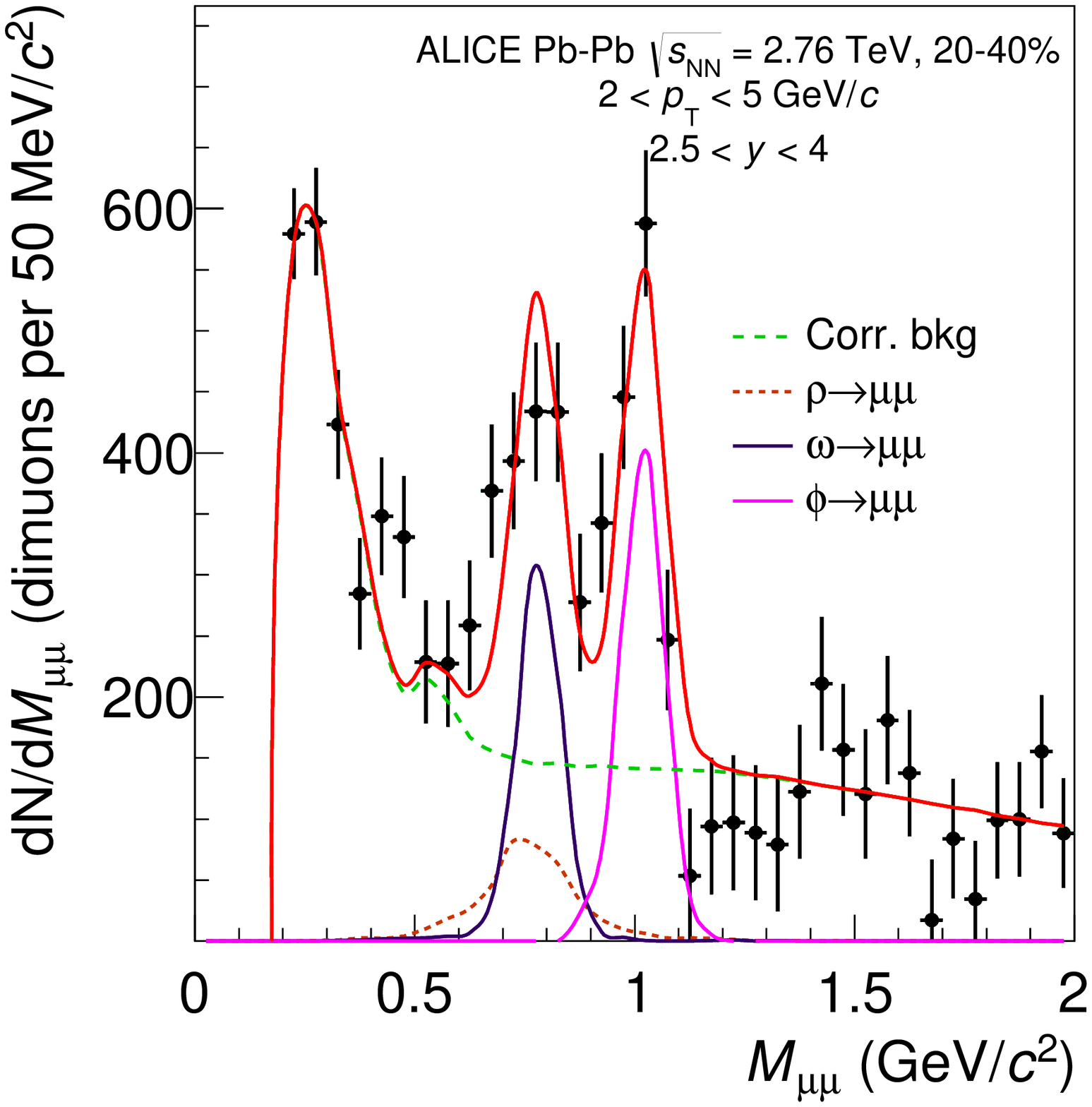}
        \includegraphics[width=0.47\textwidth]{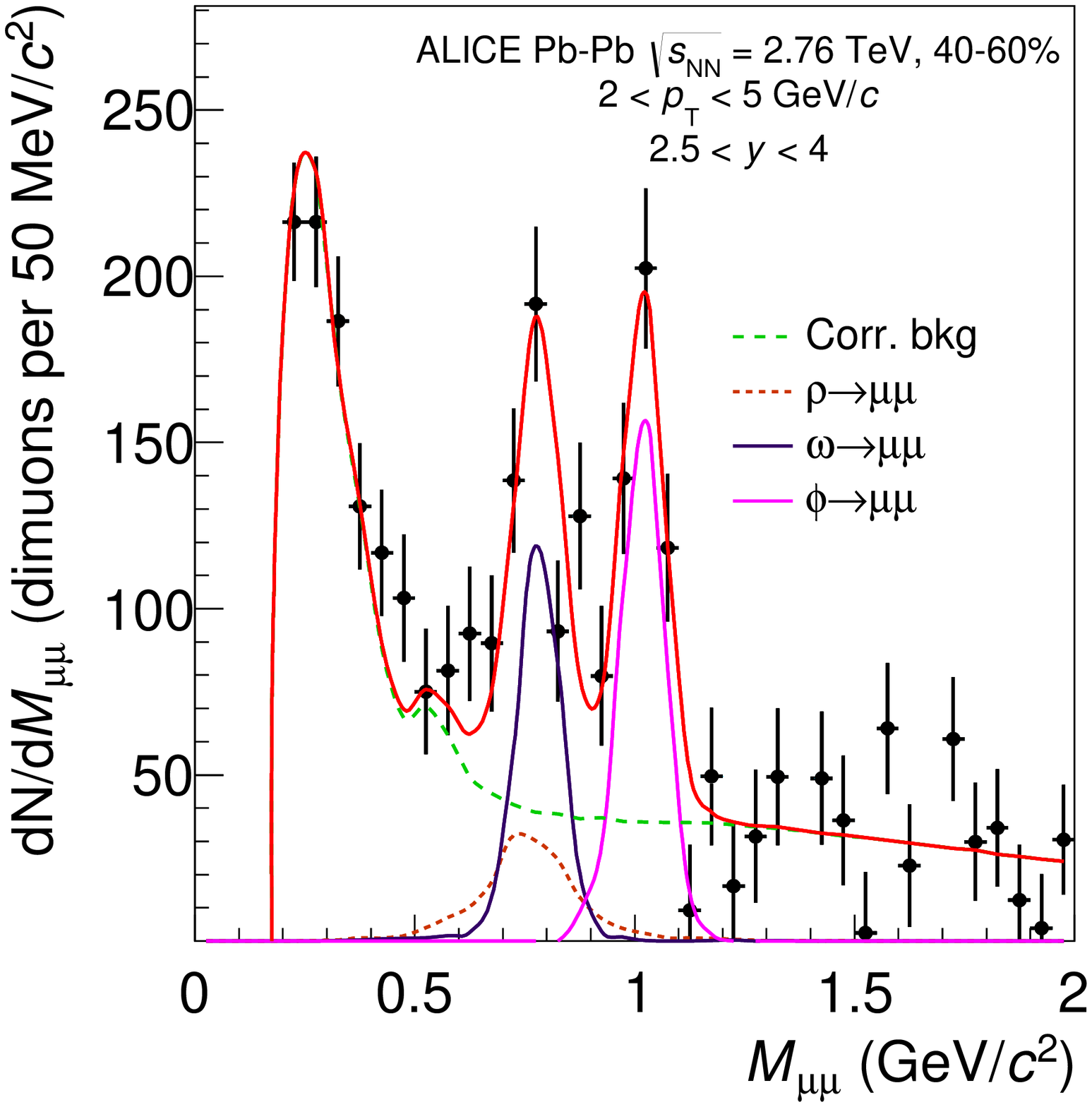}
        \includegraphics[width=0.47\textwidth]{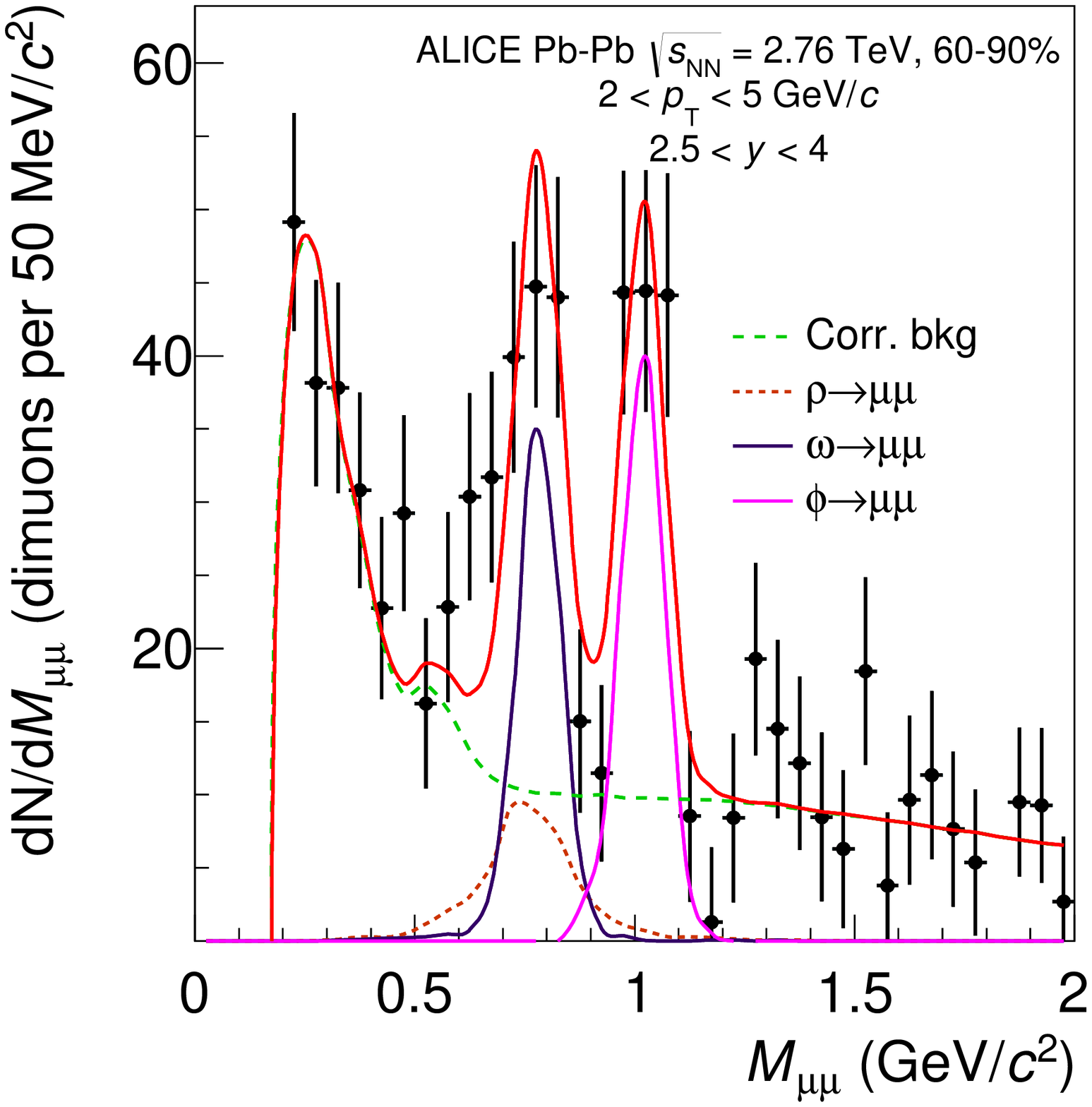}
	\caption{Invariant mass spectra in different centrality classes for 2~$< p_\mathrm{T} <$~5~GeV/\textit{c} in Pb--Pb collisions at $\sqrt{s_\mathrm{NN}}$ = 2.76~TeV. The solid red line represents the result of the fit to the hadronic cocktail; the green dashed line represents the correlated background, given by the sum of all the MC sources other than $\phi$, $\rho$ and $\omega$ mesons.}
	\label{fig:fit_centrbins}
\end{figure}

The raw invariant mass spectrum after combinatorial background subtraction is shown in Fig.~\ref{fig:fit_centrbins} 
in the four centrality classes considered in this analysis. 
The $\phi$ peak is clearly visible in all the centrality bins, superimposed to a correlated background
due to the dimuon two-body and Dalitz decays of the light resonances ($\eta$, $\eta'$, $\rho$, $\omega$) 
and the semi-muonic decays of open charm and open beauty. 
To reproduce the different processes contributing to the dimuon mass spectrum, a Monte Carlo simulation was performed using the hadronic cocktail generator first developed 
for the analysis of pp collisions at $\sqrt{s}=7$~TeV~\cite{ALICE:2011ad}. An exponential function 
\begin{equation}
\label{eq:ptdistr}
\frac{1}{\pt} \frac{\mathrm{d}N}{\mathrm{d}\pt} \propto e^{-m_\mathrm{T}/T}
\end{equation}
was used as input $\pt$ distribution of the $\phi$ meson in the generator. In this formula $m_\mathrm{T}$ is the transverse mass. 
The value of the parameter $T$ was tuned iteratively to the results from the present analysis, as shown below, with $T = (437 \pm 28)$~MeV, obtained from a fit to the $\pt$ distribution 
integrated over centrality.

The $\phi$ rapidity distribution was based on a parametrization of PYTHIA~6.4~\cite{Sjostrand:2006za}. We assume that the rapidity and $\pt$ distributions factorize.

The fit to the mass spectra obtained after the combinatorial background subtraction 
is also shown in Fig.~\ref{fig:fit_centrbins}.  
In this fit, the shape of each contribution was taken from the MC. The fit parameters allowed to vary freely were the normalizations 
of the $\eta \rightarrow \mu^+\mu^- \gamma$, $\omega \rightarrow \mu^+\mu^-$, $\phi \rightarrow \mu^+\mu^-$ and open charm contributions. 
The other processes ($\eta \rightarrow \mu^+\mu^-$, $\eta' \rightarrow \mu^+\mu^- \gamma$, $\omega \rightarrow \mu^+\mu^-\pi^0$, $\rho \rightarrow \mu^+\mu^-$ and open beauty) were fixed to the ones mentioned above, according to the relative branching ratios or cross sections, as done in~\cite{ALICE:2011ad}. In particular, the normalization of the $\rho$ relative to the $\omega$ meson was fixed requiring that $\sigma_\rho = \sigma_\omega$, as suggested both from models and pp data~\cite{ALICE:2011ad,aguilar,refId0,uras}, while the $\eta'$ contribution was derived from the $\eta$ cross section by applying the ratio of the corresponding cross sections $\sigma_{\eta'}/\sigma_{\eta}$ = 0.3 taken from the PYTHIA tunes ATLAS-CSC~\cite{2004AcPPB.35.433B} and D6T~\cite{d6t}. The ratio between the open beauty and open charm was fixed according to the results from the LHCb Collaboration in pp collisions at 7 TeV~\cite{Aaij:2010gn,
Aaij20131}.

%In particular, the normalization of the $\rho$ relative to the $\omega$ meson was fixed requiring that $\sigma_\rho = \sigma_\omega$~\cite{aguilar, refId0, uras};
%In particular, $BR (\eta \rightarrow \mu^+\mu^-)/BR(\eta \rightarrow \mu^+\mu^- \gamma) = 1.87 \cdot 10^{-2}$, $BR (\eta' \rightarrow \mu^+\mu^- \gamma)/BR(\eta \rightarrow \mu^+\mu^- \gamma) = 3.48 \cdot 10^{-1}$ , $BR(\omega \rightarrow \mu^+\mu^-\pi^0)/BR(\omega \rightarrow \mu^+\mu^-)$ = 1.44~\cite{PDG}, $\sigma_{\rho}/\sigma_{\omega} = 1$ \cite{NA27,NA60}.
Other contributions may be present in Pb--Pb collisions, such as the in-medium 
modification of the $\rho$ meson or a thermal dilepton continuum. According 
to theoretical predictions based on~\cite{ref1,vanHees2008339,sofElectroRapp}, 
the magnitude of these contributions is expected to be below the sensitivity of our measurement~\cite{CERN-LHCC-2013-014}. However, in order to take their possible
effect into account and to allow the $\phi$ signal extraction to be studied under
various hypotheses on the shape of the correlated background, two alternative empirical descriptions of the correlated background were used: the superposition of an exponential and a constant and of an exponential and a Landau distribution. In both cases, the peaks of the $\phi$ and $\rho+\omega$ were 
described with a Crystal Ball function~\cite{Gaiser:1982yw} tuned on the MC.
%The differences among the three different background descriptions constitute
The differences among these two different background descriptions and the one obtained with the hadronic cocktail constitute 
one of the main sources of the systematic uncertainty in the signal extraction. 
The width of the reconstructed $\phi$ peak is dominated by the detector resolution.  
%Due to limitations in statistics, according to the Monte Carlo simulation its value was 
%fixed to $50$~MeV, independent of centrality.
%If a fit to the mass spectrum is performed leaving the $\phi$ peak width as a free parameter, 
%the results are compatible with the MC width, with an uncertainty of about 10~MeV. 
From the MC simulation it was determined to be $\sigma_{\phi} \approx50$~MeV/$c^{2}$ 
(Gaussian width). This width was used as a fixed parameter in the fits to the 
invariant mass spectra at all centralities, in order to reduce the sensitivity 
to statistical fluctuations.
Performing the fits with the peak width as free parameter results in values compatible within uncertainties of about 10~MeV/$c^2$ with the MC result.
Likewise, if the $\phi$ peak position is left free, the result 
is compatible with its PDG value within an uncertainty of about 10~MeV/$c^2$. 
The present measurement does not allow to determine a broadening effect or a mass shift 
smaller than these uncertainties. More stringent limits are set in~\cite{Abelev:2014uua}. 
% the $\phi$ resolution does not depend on the centrality of the collision. 
%The Gaussian width of the $\phi$ peak is about 50 MeV %varies between 60 and 70 MeV/\textit{c}, depending slightly on both $\pt$ and centrality, 
%and is dominated by the detector resolution: therefore it does not allow the observation 
%of a possible broadening of the $\phi$ meson, which has been estimated to be about 2-3 times its width in vacuum~\cite{LISSAUER199115}. 
%It was verified with the Monte Carlo simulation that the $\phi$ resolution does not depend on the centrality of the collision. 

\begin{figure}[h!]
	\centering
	\includegraphics[width=0.47\textwidth]{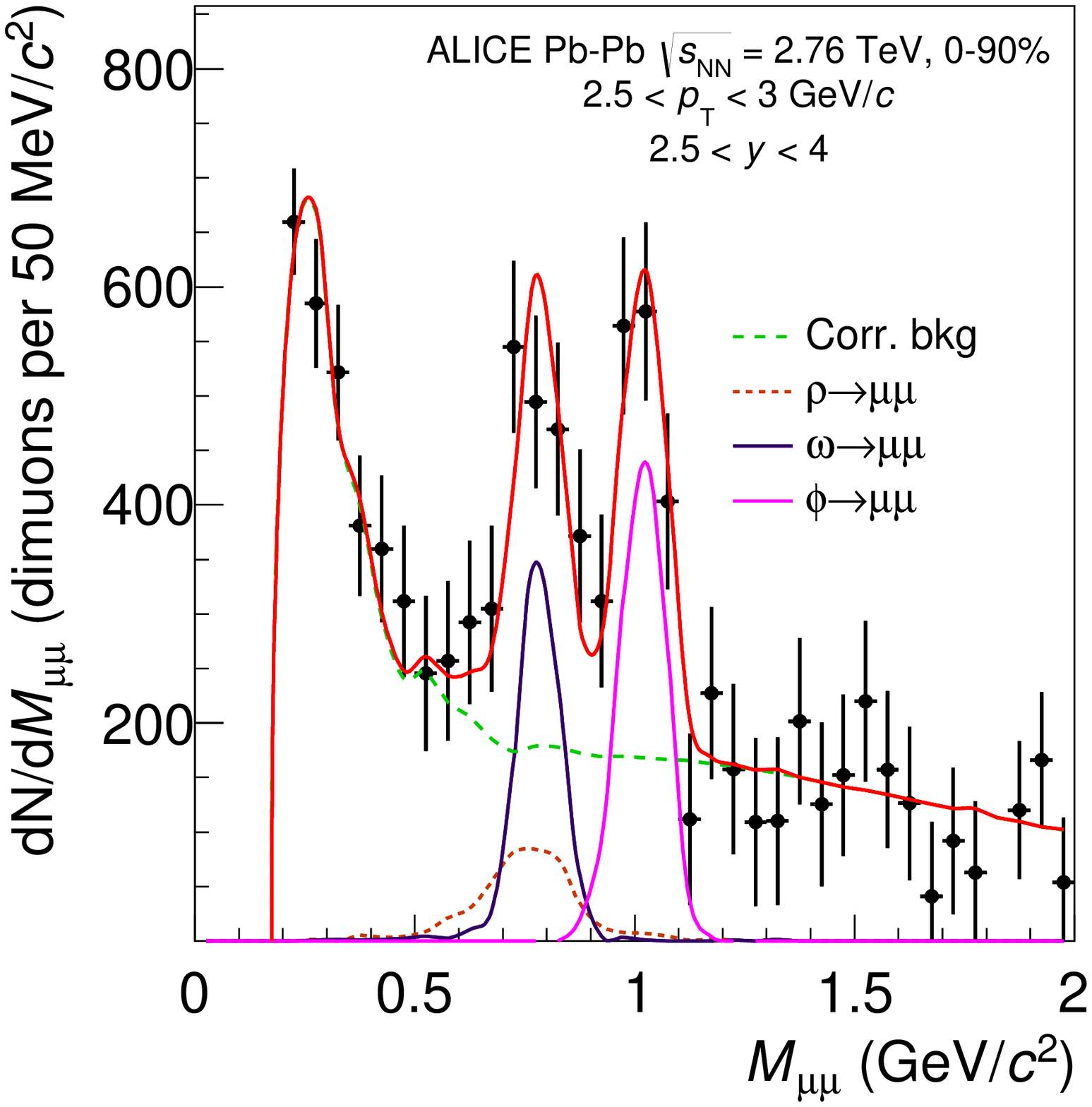}
        \includegraphics[width=0.47\textwidth]{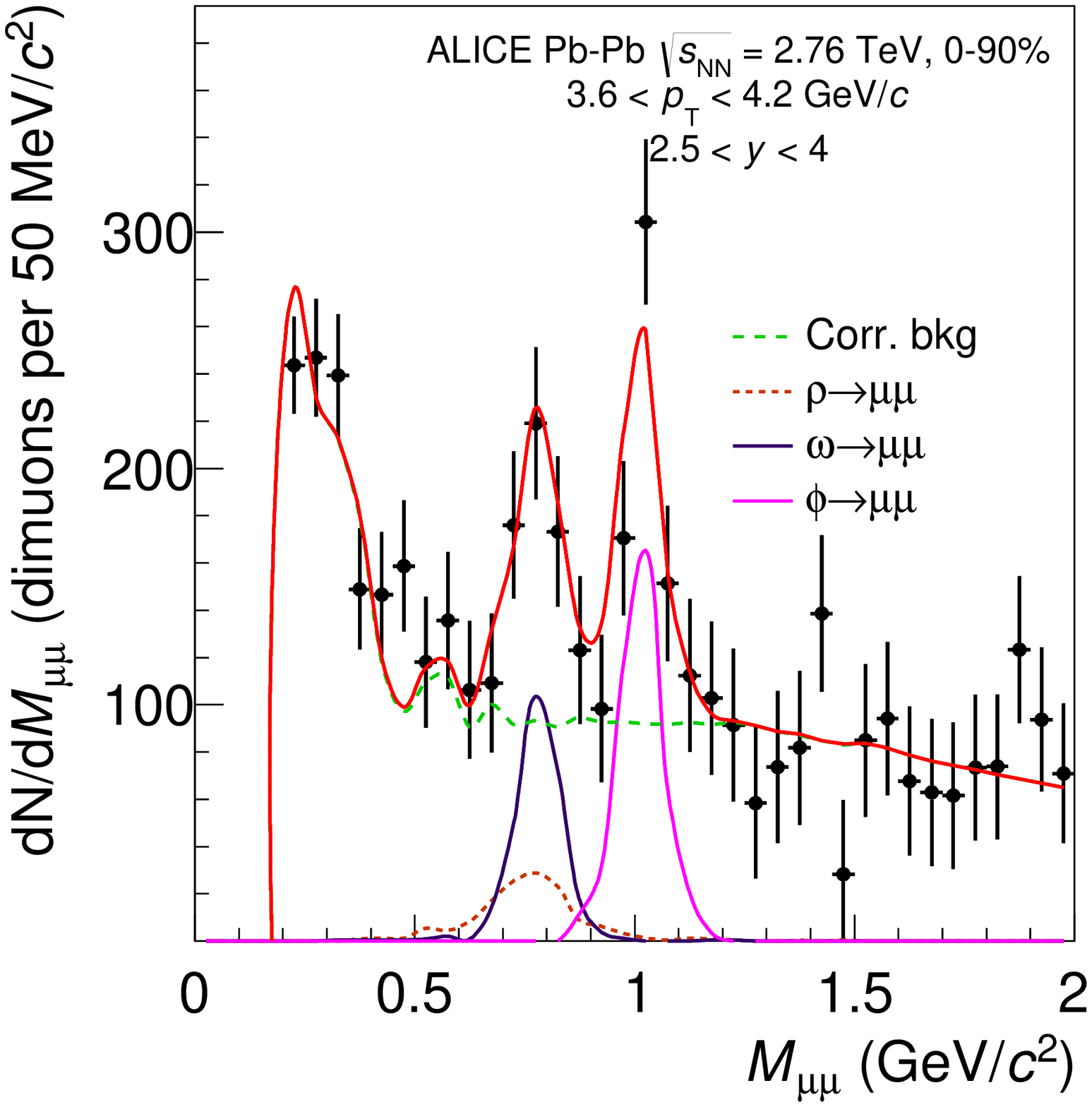}
      \caption{Invariant mass spectra for $2.5 < \pt < 3$ GeV/$c$ and $3.6 < \pt < 4.2$ GeV/$c$ in 0--90\% Pb--Pb collisions at $\sqrt{s_\mathrm{NN}}$ = 2.76~TeV. The solid red line represents the result of the fit to the hadronic cocktail; the green dashed line represents the correlated background, given by the sum of all the MC sources other than $\phi$, $\rho$ and $\omega$ mesons.}
	\label{fig:fit_ptbins}
\end{figure}

The fits of the mass spectra integrated over centrality and for 0--40\% and 40--90\% centrality classes, in different $\pt$ bins were performed as well. 
Two examples of these fits in two different $\pt$ bins 
($2.5~<~\pt~<3$~GeV/$c$ and $3.6~<~\pt~<4.2$~GeV/$c$), for 0--90\% centrality, are shown in Fig.~\ref{fig:fit_ptbins}. 

The raw number of $\phi$ mesons decaying into muon pairs $N_\phi^\mathrm{raw}$ 
and the $\phi$ yield d$N$/d$y$ in the range $2 < p_\mathrm{T} < 5$~GeV/\textit{c} are reported in Table~\ref{tab:phivsnpart} for the centrality classes considered in this analysis. Table~\ref{tab:phivspt} reports the $\phi$ yield d$^2N$/(d$y$d$\pt$) as a function of $\pt$ for 0--40\% and 40--90\% most central collisions. 
%The $\pt$ bin size was chosen according to the available statistics. 
The systematic uncertainties will be discussed below. 

\begin{table}[h!]
 \begin{center}
      \begin{tabular}{|c|c|c|}
      \hline\hline
       Centrality & $N_\phi^\mathrm{raw}$ & d$N_{\phi}$/d$y$ \\
       \hline%\hline
       0--20\% & 2337 $\pm$ 292 $\pm$ 278 & $ 0.880 \pm 0.110 \pm 0.156 $ \\
       20--40\% & 1058 $\pm$ 130 $\pm$ 86 & $ 0.387 \pm 0.048 \pm 0.060 $ \\
       40--60\% & 411 $\pm$ 51 $\pm$ 29 & $ 0.148 \pm 0.018 \pm 0.022 $ \\
       60--90\% & 105 $\pm$ 18 $\pm$ 6 & $ 0.025 \pm 0.004 \pm 0.004 $ \\
       \hline\hline
     \end{tabular}
    \caption{$N_\phi^\mathrm{raw}$ and d$N_{\phi}$/d$y$ in different centrality bins for $2 < p_\mathrm{T} < 5$~GeV/\textit{c}.}
    \label{tab:phivsnpart}
\end{center}
\end{table}

\begin{table}[ht!]
\begin{center}
\begin{tabular}{|c|c|c|c|}
\hline\hline
\multirow{2}{*} {$\pt$ (GeV/$c$)} & \multicolumn{2}{c|} {d$^2N_{\phi}$/(d$y$d$\pt$) (GeV/$c)^{-1}$} \\
& \multicolumn{1}{c} {0--40\%} & 40--90\% \\
\hline
       2--2.5 & 0.841 $\pm$ 0.185 $\pm$ 0.105 & 0.094 $\pm$ 0.021 $\pm$ 0.012 \\
       2.5--3 & 0.332 $\pm$ 0.059 $\pm$ 0.043 & 0.036 $\pm$ 0.007 $\pm$ 0.005 \\
       3--3.6 & 0.093 $\pm$ 0.016 $\pm$ 0.014 & 0.013 $\pm$ 0.002 $\pm$ 0.002 \\
       3.6--4.2 & 0.037 $\pm$ 0.007 $\pm$ 0.005 & 0.0039 $\pm$ 0.0011 $\pm$ 0.0005 \\
       4.2--5 & 0.010 $\pm$ 0.003 $\pm$ 0.002 & 0.0018 $\pm$ 0.0005 $\pm$ 0.0002 \\
\hline\hline
\end{tabular}
    \caption{$\phi$ yield d$^2N$/(d$y$d$\pt$) in different $\pt$ bins for 0--40\% and 40--90\% centrality classes.}
    \label{tab:phivspt}
\end{center}
\end{table}

The $\phi$ yield for each centrality class has been calculated as

\begin{equation}
Y_{\phi} = \frac{N_{\phi}^\mathrm{raw}}{BR_{\phi \rightarrow e^+e^-}  \cdot A \times \varepsilon \cdot  N_{MB}},
\end{equation}

where $A$ is the geometrical acceptance, $\varepsilon$ the reconstruction 
efficiency, $N_{MB}$ the number of minimum bias events in a given centrality class
and $BR_{\phi \rightarrow e^+e^-} = (2.954 \pm 0.030) \cdot 10^{-4}$ the branching 
ratio of the $\phi$ meson decay into dielectrons~\cite{Olive:2016xmw}. 
Lepton universality allows to use this value instead of the one of the dimuon decay, 
which is known with a much larger uncertainty. 
%We use this one instead of $BR_{\phi \rightarrow \mu^+\mu^-}$, because the uncertainties 
%on the latter are significantly larger and, due to lepton universality, they should be %almost identical. 

%The Monte Carlo simulations needed to correct the $N_{\phi}^{raw}$ were performed using the embedding technique, that allows to extract the reconstruction efficiency for the different sources as a function of centrality. The embedding Monte Carlo technique consists in simulating a signal particle and embedding the detector response into a real event. The embedded event is then reconstructed as if it were a normal real event. The embedding technique has the advantage of providing the most realistic background conditions. Such realistic description is necessary if the environment (high particle multiplicity) could alter the track reconstruction efficiency, as in central Pb--Pb collisions.

To estimate $A \times \varepsilon$ as a function of centrality, Monte Carlo simulations 
were performed using the embedding technique, which consists in simulating a signal 
decay and adding the corresponding simulated detector response to the raw data of a real event. 
The resulting embedded event is then reconstructed as if it were a normal real event. This technique has the 
advantage of providing the most realistic background conditions, which is necessary 
if the high particle multiplicity environment alters the track reconstruction 
efficiency, as in central Pb--Pb collisions.
The $A \times \varepsilon$ %decreases slightly as a function of centrality, 
is roughly independent from centrality, changing from 5.49\% $\pm$ 0.31\% (syst) in peripheral (60--90\%) to 5.15\% $\pm$ 0.30\% (syst) in central (0--20\%) collisions. The embedded simulations were used also to evaluate the $A \times \varepsilon$ as a function of $\pt$. The $A \times \varepsilon$ increases as a function of $\pt$ from $\sim$2.5\% for $2 < \pt < 2.5$~GeV/$c$ to $\sim$21.4\% for $4.2 < \pt < 5$~GeV/$c$.

The number of minimum bias events has been obtained from the number of opposite-sign dimuon triggers, 
scaled by the normalization factor $f_\mathrm{norm}$~\cite{Abelev:2013ila}, defined as the inverse of the probability of having a dimuon trigger in a MB event.
%as the ratio, in a MB data sample, of the number of MB events divided by the number of events fulfilling the $\mu\mu$MB trigger condition. 
Its value, averaged over the entire data sample, is $f_\mathrm{norm} = 30.56 \pm 0.01 (\mathrm{stat.}) \pm 1.10 (\mathrm{syst.})$.

% \begin{equation}
% f_\mathrm{norm} = \frac{N^\mathrm{trig}_{MB}}{N^\mathrm{trig}_{OS \& MB}} = 30.56 \pm 0.01 (\mathrm{stat.}) \pm 1.10 (\mathrm{syst.}),
% \end{equation}

%where $N^\mathrm{trig}_{MB}$ is the number of events acquired with the MB trigger and $N^{trig}_{OS \& MB}$ is the number of events triggered by both MB and dimuon triggers.

The systematic uncertainty on the $\phi$ yield was evaluated taking into account several contributions:

\begin{itemize}
\item Combinatorial background subtraction: this uncertainty was evaluated through a Monte Carlo simulation. 
The correlated muon pairs coming from the hadronic cocktail were added to the uncorrelated pairs, 
generated as described above. The relative abundance of correlated and uncorrelated muon pairs was chosen such that 
it reproduced the one in the data. The resulting mass spectrum was then subjected to the same analysis chain applied
to the data, including background subtraction with the event mixing and fit with the hadronic cocktail. The number 
of raw $\phi$ mesons obtained from the fit differs from the one actually injected in the spectrum, which is 
known {\sl a priori}. This difference was taken as an estimate of the uncertainty related to the combinatorial background subtraction. 
It decreases from 9\% in central collisions to less than 1\% in peripheral collisions, while as a function of $\pt$, it amounts to 4.8\% for 0--40\% centrality and to 1.8\% for 40--90\% centrality.

\item Shape of the correlated background: this was evaluated using two alternative empirical descriptions of the correlated background, as previously described. 
The variations of $N_\phi^\mathrm{raw}$ decrease from 6.7\% in central collisions to 2.2\% in peripheral collisions. As a function of $\pt$, it varies from 1.9\% to 8.2\% for 0--40\% centrality and from 1\% to 4\% for 40--90\% centrality. 

\item Range of the fit to the mass spectrum: three different fit ranges were tested: 0.2~$ < M_{\mu\mu} < 1.8$~GeV/$c^{2}$, $0.2< M_{\mu\mu} < 2.0$~GeV/$c^{2}$ and $0.2 < M_{\mu\mu} < 2.2$~GeV/$c^{2}$. The effect on $N_\phi^\mathrm{raw}$ is below 1\% for all centralities, except for the most peripheral bin, where it amounts to 2.1\%. As a function of $\pt$, it varies between 1.2\% and 2.9\% in central and semi-central collisions and from 1\% to 2\% in semi-peripheral and peripheral collisions.

\item Cut on single muon transverse momentum: this was evaluated by applying three different cuts, $\ptmu >$~0.7~GeV/\textit{c}, $\ptmu >$~0.85~GeV/\textit{c} and $\ptmu >$~1~GeV/\textit{c}, which lead to variations of the corresponding $A \times \varepsilon$ corrected $N_\phi^\mathrm{raw}$ ranging from 4\% to 6.3\% as a function of centrality, from 1.1\% to 10.4\% as a function of $\pt$ for 0--40\% centrality, and from 1\% to 5.7\% as a function of $\pt$ for 40--90\% centrality.
	
%\item %Systematic uncertainty related to the input $\pt$ distribution used in the simulation to calculate $A \times \varepsilon$: to evaluate this contribution, the $\pt$ distribution used as an input for the generation was fitted with two different functions, an exponential (eq.~\ref{eq:ptdistr}) and a Levy-Tsallis function~\footnote{The Levy-Tsallis distribution~\cite{tsallis} is defined as $\frac{1}{\pt}\frac{dN}{d\pt} \propto \left(1 + \frac{m_\mathrm{T} - m_{\phi}} {\beta_\mathrm{T}} \right)$, where $m_\mathrm{T}$ is the transverse mass and $\beta_\mathrm{T}$ the transverse expansion velocity of the fireball.}. The statistical uncertainty on the fit parameters was also taken into account. 
\item Systematic uncertainty of $A \times \varepsilon$: 
to evaluate this contribution, the measured $\pt$ distribution, shown in the next section, was fitted with an exponential (Eq.~\ref{eq:ptdistr}). 
The value of the $T$ parameter was used as an input to the simulation, that was then repeated 
varying $T$ by one standard deviation $\sigma_T$. %according to its uncertainty.
The half of the difference between the $A \times \varepsilon$ values obtained using $T\pm \sigma_T$ as input parameter was taken as an estimation of its systematic uncertainty.
As a function of centrality, it amounts to about 5.7\% with no significant dependence on the collision centrality; it is $<1$\% as a function of $\pt$.

\item Tracking and trigger efficiencies: the corresponding systematic uncertainties were determined from data and MC simulations as detailed in~\cite{Abelev:2013ila}. % and amount to 11\% and 2\%, respectively as a function of centrality. 
They are correlated as a function of centrality, amounting respectively to 11\% and 2\%, and uncorrelated as a function of $\pt$, varying in this case of 8--14\% and 2--4\% respectively.
	
\item Matching efficiency: the uncertainty on the matching efficiency between the tracks reconstructed in the tracking chambers and the ones reconstructed in the trigger chambers
amounts to 1\%. It is correlated as a function of centrality and uncorrelated as a function of $\pt$~\cite{Abelev:2013ila}.

\item Centrality limits: the effects of the uncertainty on the value of the V0 signal amplitude corresponding to 90\% of the hadronic Pb--Pb cross section were estimated by varying such a value by $\pm$0.5\%~\cite{Abelev:2013qoq} and redefining correspondingly the centrality intervals. The systematic effect of $N_\phi^\mathrm{raw}$ is negligible in all centrality bins, except for the most peripheral one, where it amounts to 3\%. It is correlated as a function of $\pt$, amounting to less than 1\%.

\item Uncertainty of the $\phi$ branching ratio into dielectrons ($\sim$1\%)~\cite{Olive:2016xmw}, correlated as a function of $\pt$ and centrality.

\item Uncertainty on $f_\mathrm{norm}$ ($\sim$3.6\%), correlated as a function of $\pt$ and centrality~\cite{Abelev:2013ila}.
\end{itemize}

\begin{table}[t!]
\begin{center}
%\begin{footnotesize}
     \begin{tabular}{|c|c|c|c|}
     %\toprule
     \hline\hline
     Source & vs centrality & vs $\pt$ & vs $\pt$\\
     & (2 $< \pt <$ 5 GeV/$c$) & (0--40\%) & (40--90\%) \\
     \hline
      Combinatorial background subtraction & 0.6--9.0\% & 4.8\% & 1.8\% \\
      Correlated background shape & 2.2--6.7\% & 1.9--8.2\% & 1.0--4.0\% \\
      Fit range & 0.4--2.1\% & 1.2--2.9\% & 1.0--2.0\% \\
      Cut on $\ptmu$ & 4.0--6.7\% & 1.1--10.4\% & 1.0--5.7\% \\
      $A \times \varepsilon$ ($\phi$) & 5.5--5.9\% & $<1$\% & $<1$\% \\
      Tracking efficiency & 11\%$^{*}$ & 8--14\% & 8--14\% \\
      Trigger efficiency & 2\%$^{*}$ & 2--4\% & 2--4\% \\
      Matching efficiency & 1\%$^{*}$ & 1\% & 1\% \\
      Centrality limits & 0-3\% & $<1\%^{*}$ & $<1\%^{*}$ \\
      $BR_{\phi \rightarrow e^+e^-}$ & 1\%$^{*}$ & 1\%$^{*}$ & 1\%$^{*}$ \\
      $f_\mathrm{norm}$ & 3.6\%$^{*}$ & 3.6\%$^{*}$ & 3.6\%$^{*}$ \\
      \hline\hline
      \end{tabular}
%\end{footnotesize}
\caption{Systematic uncertainties on $\phi$ yield, for $2.5 < y <4$; the correlated uncertainties are marked with an $^{*}$.}	
\label{tab:phi_yield_syst}
\end{center}
\end{table}

These values are summarized in Table~\ref{tab:phi_yield_syst}.

\section{Results}
\label{sect:Results}

\begin{figure}[bt!]
	\centering
	\includegraphics[width=0.65\textwidth]{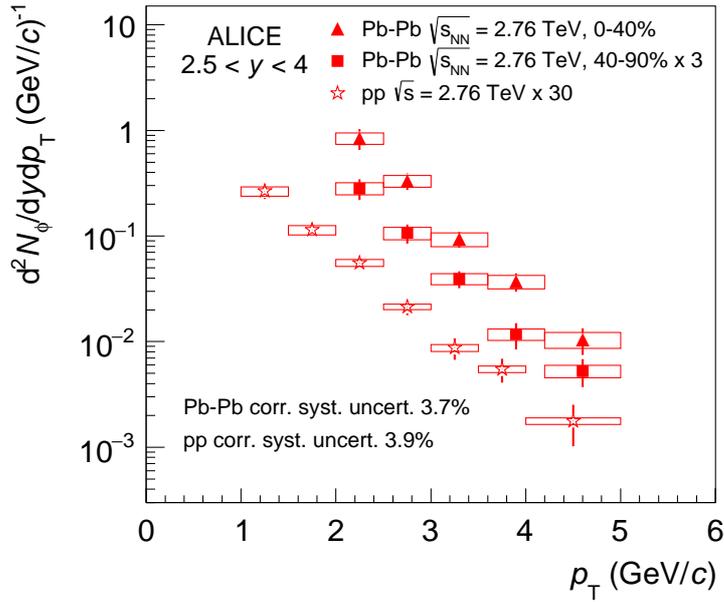}
	\caption{$\phi$ yield as a function of $\pt$ at forward rapidity in pp~\cite{Adam:2015jca} and Pb--Pb collisions for different centralities.
	The distributions have been scaled differently for better visibility.}
	\label{fig:phi_yield_PbPb_pp}
\end{figure}

Figure~\ref{fig:phi_yield_PbPb_pp} shows the $\pt$ spectra in Pb--Pb collisions in the centrality ranges 0--40\% and 40--90\%. The pp spectrum~\cite{Adam:2015jca} is also reported for comparison. 
The $\pt$ distribution in Pb--Pb collisions is softer than in pp in the measured transverse momentum range.  

\begin{figure}[t!]
	\centering
	\includegraphics[width=0.99\textwidth]{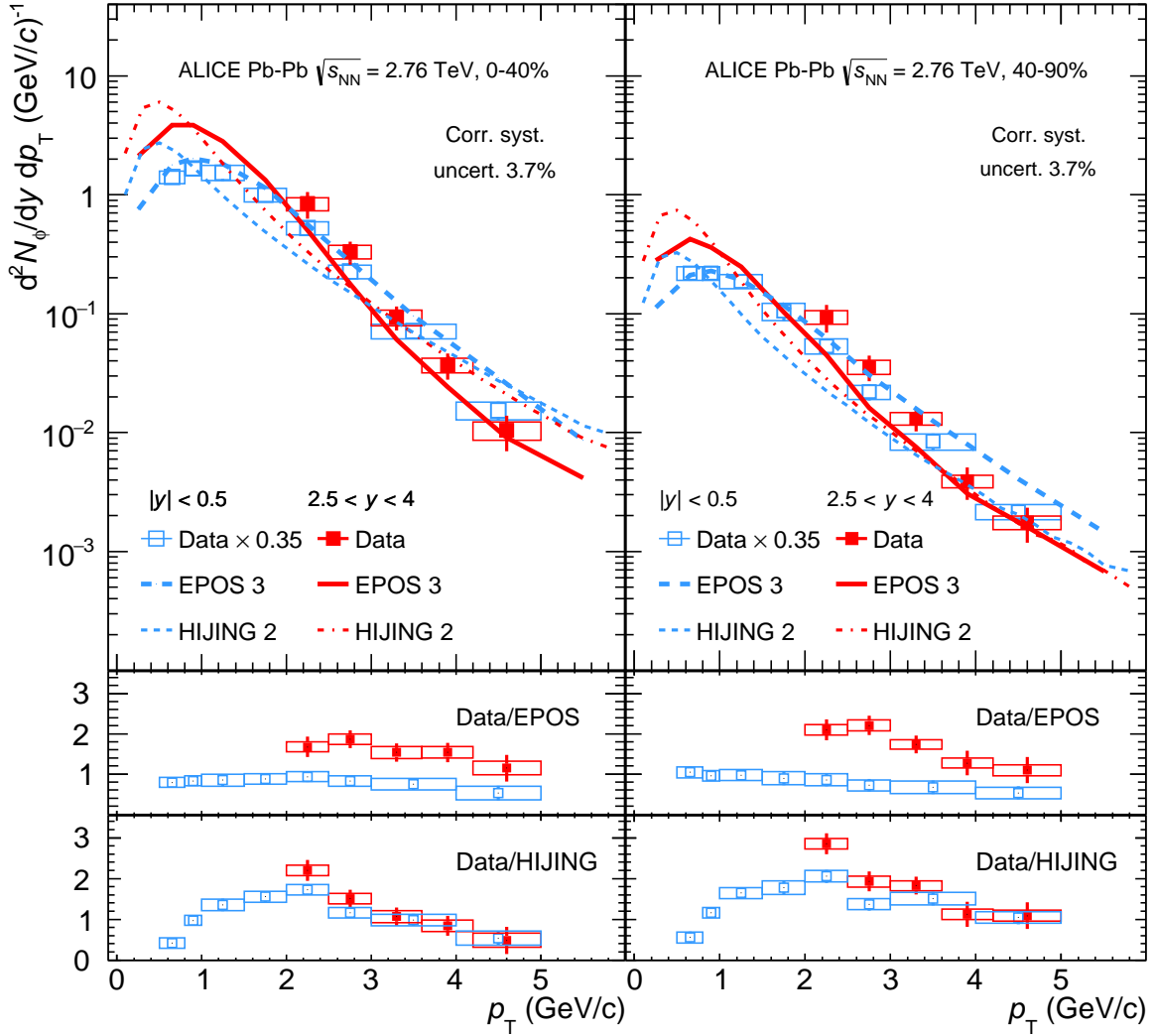}
	\caption{Top panel: comparison between the $\phi$ yield as a function of $\pt$ with the EPOS~3.101~\cite{Werner:2012xh, Drescher200193, PhysRevC.89.064903} 
	and HIJING~2.0~\cite{PhysRevD.44.3501} event generators, at forward and midrapidity~\cite{Abelev:2014uua}, for 0-40\% (left) and 40-90\% centrality (right). 
	The same scale factors applied to data were also used for the models. The transparent boxes represent the uncorrelated systematic uncertainties 
	at forward rapidity and the total systematic uncertainties (including correlated and uncorrelated components) at midrapidity. 
	Lower panels: ratios between the measured yields and the calculations by EPOS and HIJING.}
	\label{fig:phi_yield_pt}
\end{figure}

In Fig.~\ref{fig:phi_yield_pt} the $\pt$ spectra are compared with the EPOS~3.101\footnote{We used a version of the EPOS~3.101 
generator, customized by the authors, in which the spectra 
for the $\phi$ decaying into dimuons were determined by 
the kinematics of the $\phi$ at the moment of its decay, assuming that the decay
muons do not interact with the surrounding medium. On the other side, 
kaons originating from the $\phi$ decay are allowed to rescatter inside the 
hadronic medium and thus emerge with an altered momentum distribution.} event generator~\cite{Werner:2012xh, Drescher200193, PhysRevC.89.064903}, which utilizes a core--corona approach in which the core component undergoes a hydrodynamic expansion, 
%that implements the hydrodynamic evolution of the system, 
and the HIJING~2.0 model~\cite{PhysRevD.44.3501}, 
which does not include hydrodynamic effects in the calculation. 
The results obtained at midrapidity~\cite{Abelev:2014uua} are also shown. 
EPOS fairly reproduces the data at midrapidity for all centralities, 
although the $\pt$ spectra are slightly harder than the measured ones.
At forward rapidity, the calculation underestimates the $\pt$ spectra at all centralities, 
approaching the data only at $\pt \sim$4~GeV/$c$. 
% aggiunta di chinellato
It has to be noted that this disagreement is significantly worsened if the core component in EPOS 
were to be switched off. Qualitatively, this suggests that the steepness of the forward $\phi$ spectra 
in Pb--Pb is a consequence of the interplay between radial flow at low- to mid-$\pt$, 
which increases the $\phi$ yield in the lowest measured $\pt$ range, 
and a relatively unchanged contribution from hard processes at higher transverse momenta. 
HIJING underestimates the data and shows a harder $\pt$ shape at both mid- and 
forward rapidity. In particular, at forward rapidity, the disagreement with the
data on the shape of the $\pt$ distribution is stronger than for EPOS. 

\begin{figure}[tb]
	\centering
		\includegraphics[width=0.65\textwidth]{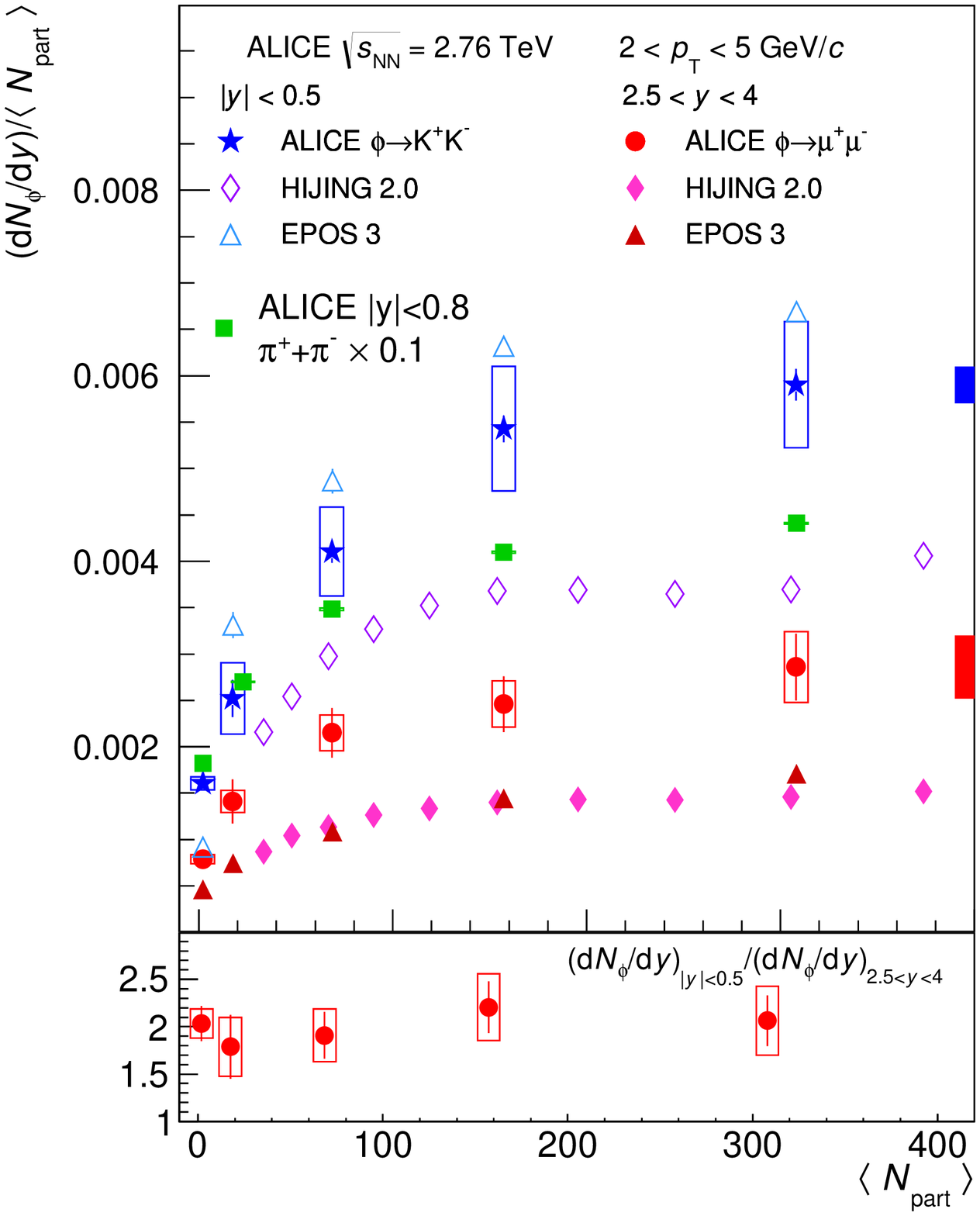}
		\caption{Top panel: comparison between $(\mathrm{d}N_{\phi}/\mathrm{d}y)/\npart$ as a 
		function of $\npart$ measured in the muon decay channel at forward rapidity 
        and in the kaon decay channel at midrapidity, in Pb--Pb collisions 
        at $\sqrt{s_\mathrm{NN}}= 2.76$~TeV, for $2 < p_\mathrm{T} <5$~GeV/\textit{c}. 
        The corresponding points in pp collisions at $\npart = 2$ are also shown. 
        Transparent boxes represent the uncorrelated systematic uncertainties at forward rapidity and the total systematic uncertainties (including correlated and uncorrelated components) at midrapidity.
        The shaded red box represents the correlated systematic uncertainties at forward rapidity, the shaded blue box represents the normalization uncertainty at midrapidity. 
        Results from the EPOS~3.101 and HIJING~2.0 event generators are shown for comparison.
        The rapidity density per participant for pions at midrapidity is also reported, 
        scaled to 0.1. 
        Bottom panel: ratio between $\phi$ rapidity densities per participant at mid- and forward rapidity, 
        in pp (open circle) and Pb--Pb collisions (full circles).}
  	\label{fig:phi_yield_mumu_KK}
\end{figure}

Figure~\ref{fig:phi_yield_mumu_KK} shows the $\phi$ rapidity density per participant 
% in the same $\pt$ range 
as a function of $\npart$. 
The result in pp collisions at the same energy~\cite{Adam:2015jca} is also shown: 
the $\phi$ yield per participant already grows by a factor of about 1.8
when going from pp to peripheral Pb--Pb collisions. This factor increases to 
about 4 when going from pp to central Pb--Pb collisions. No sizeable dependence on rapidity is observed. 
The ratio between the rapidity densities at mid- and forward 
rapidity is $\sim 2$, both in pp and Pb--Pb collisions, where it is roughly constant as a function of centrality, 
as shown in the lower panel of the same figure. 

The rapidity density per participant is also plotted in Fig.~\ref{fig:phi_yield_mumu_KK}
for pions at midrapidity~\cite{Adam:2015kca}. The rapidity density increases from pp to Pb--Pb collisions 
faster for the $\phi$ than for pions. The increase of the $\phi/\pi$ ratio in the intermediate $\pt$ region 
is interpreted in terms of radial flow, whose magnitude grows as a function of the collision centrality. 
The similar increase of the $\phi$ at mid- and forward rapidity suggests that the magnitude
of radial flow is similar in the two rapidity regions considered. 

The comparison with HIJING and EPOS at forward rapidity shows that both calculations 
predict a similar evolution of the yield with the collision centrality.
In this rapidity region, both models underestimate the yield by about a factor of two, independently of centrality.
Different results are obtained at midrapidity, where HIJING largely underestimates the yield, 
while EPOS qualitatively reproduces the trend as a function of $\npart$, even though it overestimates 
the data by about 30\% in peripheral collisions and 13\% in central collisions. 
%The ratio between the rapidity densities at mid- and forward rapidity is independent from centrality 
%also for EPOS and HIJING. %(see again bottom panel of Fig.~\ref{fig:phi_yield_mumu_KK}). 

\begin{figure}[t]
%\begin{minipage}{0.49 \textwidth}
\centering
\includegraphics[width=0.65 \textwidth]{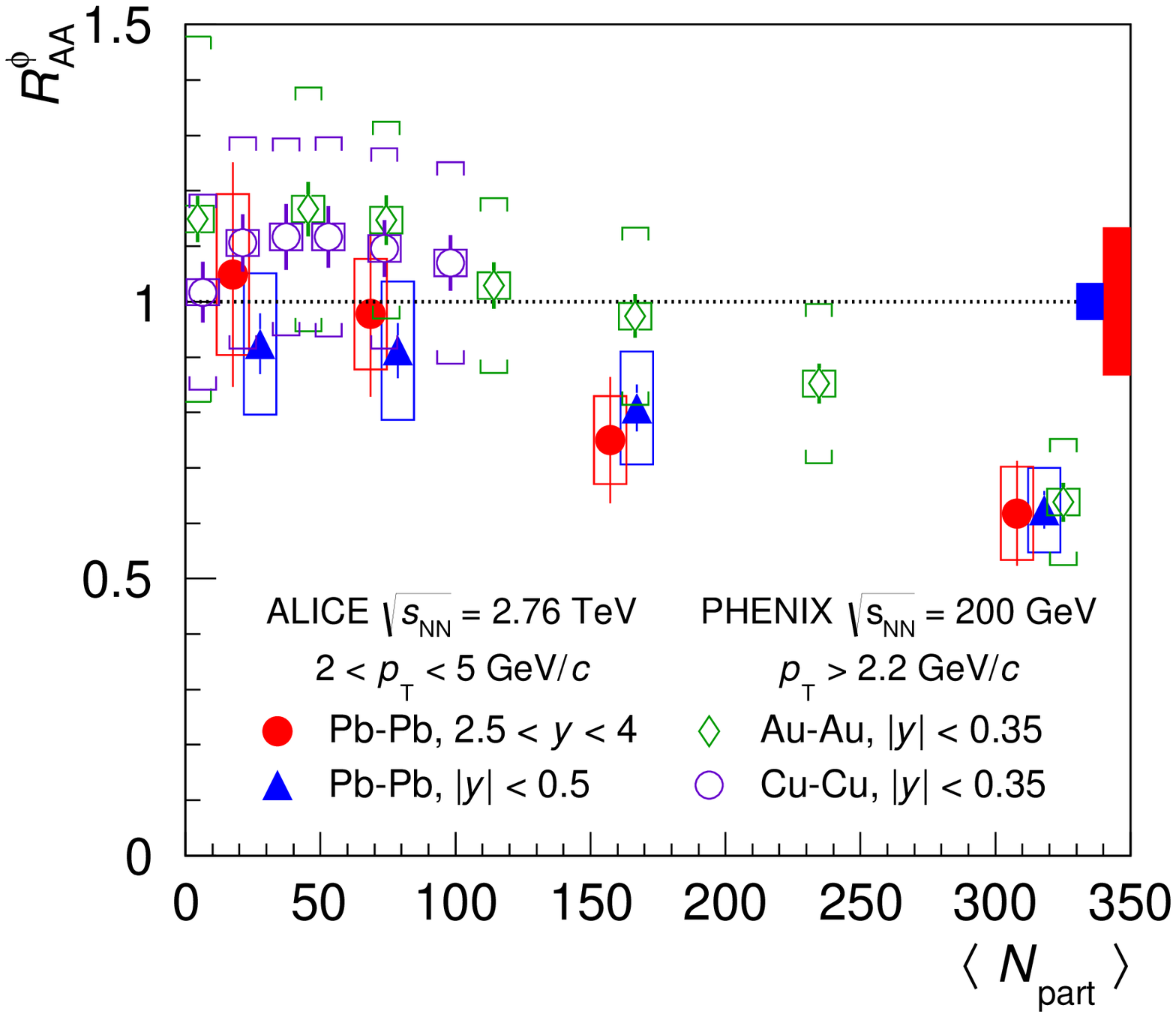}
\caption{\label{fig:raa} $R_{\mathrm{AA}}$ of the $\phi$  meson as a function of $\langle N_{\mathrm{part}} \rangle$ for $2.5<y<4$, compared with the 
	ALICE measurement for $|y|<0.5$. All the midrapidity points have been displaced by $\npart$ = 10 for better visibility. 
	Transparent boxes represent the uncorrelated systematic uncertainties at forward rapidity and the total systematic uncertainties (including correlated and uncorrelated components) at midrapidity.
        The shaded red box represents the correlated systematic uncertainties at forward rapidity, the shaded blue box represents the normalization uncertainty at midrapidity. Results from PHENIX in Au-Au and Cu-Cu collisions at $\sqrt{s_\mathrm{NN}}$ = 200 GeV are also shown for comparison.}
%\end{minipage} 
\end{figure}

The nuclear modification factor is defined as the yield ratio of nucleus--nucleus collisions to inelastic pp
collisions, scaled with the average nuclear overlap function $\left\langle  T_\mathrm{AA} \right\rangle$. 
For a given centrality and integrated over the considered $\pt$ and $y$ ranges, it is obtained as 
\begin{equation}
R_\mathrm{AA} = {\frac{\mathrm{d}N/\mathrm{d}y}{\mathrm{d} \sigma_\mathrm{pp} /\mathrm{d}y \left\langle  T_\mathrm{AA} \right\rangle}},
\end{equation}
where $\mathrm{d}N/\mathrm{d}y$ is the $\phi$ rapidity density and 
$\mathrm{d} \sigma_\mathrm{pp} /\mathrm{d}y= 113 \pm 10 \mathrm{(stat)} \pm 7 \mathrm{(syst)}~\mu$b~\cite{Adam:2015jca} 
is the $\phi$ production cross section in pp collisions at the same energy, integrated over the corresponding $\pt$ and rapidity range.

Figure~\ref{fig:raa} shows the $R_\mathrm{AA}$ measured as a function of the average number of participants. 
% Transparent boxes represent the uncorrelated systematic uncertainties at forward rapidity and the total systematic uncertainties (including correlated and uncorrelated components) at midrapidity. The shaded red box represents the correlated systematic uncertainties at forward rapidity, the shaded blue box represents the normalization uncertainty at midrapidity.
The systematic uncertainties at forward rapidity are summarized in Table~\ref{tab:raa_syst}.
\begin{table}[h!]
\begin{center}
%\begin{footnotesize}
     \begin{tabular}{|c|c|}
     \hline\hline%
     Source & Systematic uncertainty \\
     \hline
      $N_{\phi}^{raw}$ & 5.9--11.9\% \\
      $A \times \varepsilon$ ($\phi$) & 5.5--5.9\% \\
      $T_\mathrm{AA}$ & 3.2--9.7\% \\
      Centrality limits & 0-3\% \\
      Tracking efficiency & 11\%$^{*}$ \\
      Trigger efficiency & 2\%$^{*}$\\
      Matching efficiency & 1\%$^{*}$ \\
      $f_\mathrm{norm}$ & 3.6\%$^{*}$ \\
      $\sigma_\mathrm{pp}$ & 7.2\%$^{*}$ \\
      \hline\hline
      \end{tabular}
%\end{footnotesize}
\caption{Systematic uncertainties for $R_{\mathrm{AA}}$ as a function of $\npart$, for $2 < \pt < 5$ GeV/\textit{c}; the correlated uncertainties are marked with an $^{*}$.}
\label{tab:raa_syst}
\end{center}
\end{table}
In peripheral collisions, the nuclear modification factor is compatible with unity
within uncertainties, indicating that these collisions behave as a superposition of incoherent pp collisions. 
In most central collisions, $R_\mathrm{AA}$ at forward rapidity is reduced to about 0.65, 
showing a clear suppression of the $\phi$ multiplicity with respect to the pp
reference in the intermediate $\pt$ region. A qualitatively similar behaviour
was observed also by the PHENIX experiment in Au-Au collisions 
at $\sqrt{s_\mathrm{NN}}=200$~GeV for $\pt>2.2$~GeV/$c$ 
at midrapidity~\cite{Adare:2010pt}.   

The comparison with the ALICE results obtained at midrapidity shows that 
the two data sets agree within the present uncertainties. 
%The independence of the $R_\mathrm{AA}$ as a function of $\npart$ 
%from the rapidity of the detected particle has also been observed 
%for the $J/\psi$ meson~\cite{Abelev:2013ila}. 

\section{Conclusions}
\label{sect:Conclusions}

$\phi$ meson production was measured via its dimuon decay channel 
in Pb--Pb collisions at $\sqrt{s_\mathrm{NN}} = 2.76$~TeV
at forward rapidity. For intermediate transverse momenta ($2<\pt<5$~GeV/$c$) 
the $\pt$ spectra in Pb--Pb collisions are softer than in pp.
%This observation hints for the presence of radial flow in Pb--Pb collisions at forward rapidity. 

The yield per participant increases with the collision centrality, 
similarly to the yield measured in the kaon decay channel at midrapidity.
The ratio between the yields in the two rapidity regions is constant as a 
function of centrality. The rapidity density increases from pp to Pb--Pb collisions 
faster for the $\phi$ than for pions, suggesting the presence of radial flow, 
whose effect increases with the collision centrality with similar magnitudes
at forward and midrapidity.  

The measured yields as a function of $\pt$ or centrality were compared with results 
from the EPOS event generator and the HIJING Monte Carlo model.
The two calculations predict similar centrality dependencies at forward
rapidity, underestimating the measured yield by a factor of $\sim$2 for all centralities.
At midrapidity, EPOS qualitatively reproduces 
the trend as a function of the collision centrality, while
HIJING largely underestimates the yield. 
Regarding the shape of the $\pt$ spectra, EPOS correctly reproduces the data at midrapidity, 
while it predicts harder transverse momentum distributions at forward rapidity. 
HIJING predicts harder $\pt$ distributions at both mid- and forward 
rapidity. 
%{\bf The comparison with the models would suggest that EPOS underestimates the radial flow at forward rapidity, originated from the hydrodynamical expansion of the bulk in Pb--Pb collisions. However, the interpretation of this comparison is at the moment not conclusive, due to the fact that EPOS predicts $\pt$ spectra harder than data already in pp collisions, the baseline to identify hydrodynamical effects in Pb--Pb. A fine tuning of the EPOS model in pp collisions is thus needed, to clarify whether the observed differences between pp and Pb--Pb data can be actually ascribed to radial flow.}

The integrated nuclear modification factor, measured as a function of $\npart$, is 
compatible with unity for peripheral and semi-peripheral collisions, while
in most central collisions it is reduced to about 0.65. 
The results at forward rapidity are in 
agreement within the uncertainties with the ones at midrapidity.
The similarity of the two results hints for similar mechanisms driving
the interaction of the $\phi$ meson with the bulk and its hydrodynamical evolution, 
in the two rapidity ranges at intermediate $\pt$.

%%%%% acknowledgements
\newenvironment{acknowledgement}{\relax}{\relax}
\begin{acknowledgement}
\section*{Acknowledgements}
% Version: 2017-12-21

The ALICE Collaboration would like to thank all its engineers and technicians for their invaluable contributions to the construction of the experiment and the CERN accelerator teams for the outstanding performance of the LHC complex.
The ALICE Collaboration gratefully acknowledges the resources and support provided by all Grid centres and the Worldwide LHC Computing Grid (WLCG) collaboration.
The ALICE Collaboration acknowledges the following funding agencies for their support in building and running the ALICE detector:
A. I. Alikhanyan National Science Laboratory (Yerevan Physics Institute) Foundation (ANSL), State Committee of Science and World Federation of Scientists (WFS), Armenia;
Austrian Academy of Sciences and Nationalstiftung f\"{u}r Forschung, Technologie und Entwicklung, Austria;
Ministry of Communications and High Technologies, National Nuclear Research Center, Azerbaijan;
Conselho Nacional de Desenvolvimento Cient\'{\i}fico e Tecnol\'{o}gico (CNPq), Universidade Federal do Rio Grande do Sul (UFRGS), Financiadora de Estudos e Projetos (Finep) and Funda\c{c}\~{a}o de Amparo \`{a} Pesquisa do Estado de S\~{a}o Paulo (FAPESP), Brazil;
Ministry of Science \& Technology of China (MSTC), National Natural Science Foundation of China (NSFC) and Ministry of Education of China (MOEC) , China;
Ministry of Science and Education, Croatia;
Ministry of Education, Youth and Sports of the Czech Republic, Czech Republic;
The Danish Council for Independent Research | Natural Sciences, the Carlsberg Foundation and Danish National Research Foundation (DNRF), Denmark;
Helsinki Institute of Physics (HIP), Finland;
Commissariat \`{a} l'Energie Atomique (CEA) and Institut National de Physique Nucl\'{e}aire et de Physique des Particules (IN2P3) and Centre National de la Recherche Scientifique (CNRS), France;
Bundesministerium f\"{u}r Bildung, Wissenschaft, Forschung und Technologie (BMBF) and GSI Helmholtzzentrum f\"{u}r Schwerionenforschung GmbH, Germany;
General Secretariat for Research and Technology, Ministry of Education, Research and Religions, Greece;
National Research, Development and Innovation Office, Hungary;
Department of Atomic Energy Government of India (DAE), Department of Science and Technology, Government of India (DST), University Grants Commission, Government of India (UGC) and Council of Scientific and Industrial Research (CSIR), India;
Indonesian Institute of Science, Indonesia;
Centro Fermi - Museo Storico della Fisica e Centro Studi e Ricerche Enrico Fermi and Istituto Nazionale di Fisica Nucleare (INFN), Italy;
Institute for Innovative Science and Technology , Nagasaki Institute of Applied Science (IIST), Japan Society for the Promotion of Science (JSPS) KAKENHI and Japanese Ministry of Education, Culture, Sports, Science and Technology (MEXT), Japan;
Consejo Nacional de Ciencia (CONACYT) y Tecnolog\'{i}a, through Fondo de Cooperaci\'{o}n Internacional en Ciencia y Tecnolog\'{i}a (FONCICYT) and Direcci\'{o}n General de Asuntos del Personal Academico (DGAPA), Mexico;
Nederlandse Organisatie voor Wetenschappelijk Onderzoek (NWO), Netherlands;
The Research Council of Norway, Norway;
Commission on Science and Technology for Sustainable Development in the South (COMSATS), Pakistan;
Pontificia Universidad Cat\'{o}lica del Per\'{u}, Peru;
Ministry of Science and Higher Education and National Science Centre, Poland;
Korea Institute of Science and Technology Information and National Research Foundation of Korea (NRF), Republic of Korea;
Ministry of Education and Scientific Research, Institute of Atomic Physics and Romanian National Agency for Science, Technology and Innovation, Romania;
Joint Institute for Nuclear Research (JINR), Ministry of Education and Science of the Russian Federation and National Research Centre Kurchatov Institute, Russia;
Ministry of Education, Science, Research and Sport of the Slovak Republic, Slovakia;
National Research Foundation of South Africa, South Africa;
Centro de Aplicaciones Tecnol\'{o}gicas y Desarrollo Nuclear (CEADEN), Cubaenerg\'{\i}a, Cuba and Centro de Investigaciones Energ\'{e}ticas, Medioambientales y Tecnol\'{o}gicas (CIEMAT), Spain;
Swedish Research Council (VR) and Knut \& Alice Wallenberg Foundation (KAW), Sweden;
European Organization for Nuclear Research, Switzerland;
National Science and Technology Development Agency (NSDTA), Suranaree University of Technology (SUT) and Office of the Higher Education Commission under NRU project of Thailand, Thailand;
Turkish Atomic Energy Agency (TAEK), Turkey;
National Academy of  Sciences of Ukraine, Ukraine;
Science and Technology Facilities Council (STFC), United Kingdom;
National Science Foundation of the United States of America (NSF) and United States Department of Energy, Office of Nuclear Physics (DOE NP), United States of America.
    %%%%%%% done by webmaster team
\end{acknowledgement}

%%%%%%%% Bibliography (In case of using bibtex generate the bbl requested by arXiv)
\bibliographystyle{utphys}   % Remember we use title in the biblio
% \bibliography{biblio}
% \input {bibliography.bib}  

%\bibliographystyle{unsrt}
\bibliography{bibliography}

%%%%%%%%% appendix with author list
\newpage
\appendix
\section{The ALICE Collaboration}
\label{app:collab}
% Collaboration: CERN-LHC-ALICE
% Generation Date is 2017-Dec-21

% How to use:
%%%%%%%%% appendix with author list
%\appendix
%\section{The ALICE Collaboration}
%\label{app:collab}
%\input{Alice_Authorslist_XXXX-Axx-XX.tex}
\begingroup
\small
\begin{flushleft}
S.~Acharya\Irefn{org138}\And 
F.T.-.~Acosta\Irefn{org22}\And 
D.~Adamov\'{a}\Irefn{org94}\And 
J.~Adolfsson\Irefn{org81}\And 
M.M.~Aggarwal\Irefn{org98}\And 
G.~Aglieri Rinella\Irefn{org36}\And 
M.~Agnello\Irefn{org33}\And 
N.~Agrawal\Irefn{org48}\And 
Z.~Ahammed\Irefn{org138}\And 
S.U.~Ahn\Irefn{org77}\And 
S.~Aiola\Irefn{org143}
% ...
\And
\renewcommand\labelenumi{\textsuperscript{\theenumi}~}

\section*{Affiliation notes}
\renewcommand\theenumi{\roman{enumi}}
\begin{Authlist}
\item \Adef{org*}Deceased
\item \Adef{orgI}Dipartimento DET del Politecnico di Torino, Turin, Italy
\item \Adef{orgII}M.V. Lomonosov Moscow State University, D.V. Skobeltsyn Institute of Nuclear, Physics, Moscow, Russia
\item \Adef{orgIII}Department of Applied Physics, Aligarh Muslim University, Aligarh, India
\item \Adef{orgIV}Institute of Theoretical Physics, University of Wroclaw, Poland
\end{Authlist}

\section*{Collaboration Institutes}
\renewcommand\theenumi{\arabic{enumi}~}
\begin{Authlist}
\item \Idef{org1}A.I. Alikhanyan National Science Laboratory (Yerevan Physics Institute) Foundation, Yerevan, Armenia
\item \Idef{org2}Benem\'{e}rita Universidad Aut\'{o}noma de Puebla, Puebla, Mexico
\item \Idef{org3}Bogolyubov Institute for Theoretical Physics, National Academy of Sciences of Ukraine, Kiev, Ukraine
\item \Idef{org4}Bose Institute, Department of Physics  and Centre for Astroparticle Physics and Space Science (CAPSS), Kolkata, India
\item \Idef{org5}Budker Institute for Nuclear Physics, Novosibirsk, Russia
\item \Idef{org6}California Polytechnic State University, San Luis Obispo, California, United States
\item \Idef{org7}Central China Normal University, Wuhan, China
\item \Idef{org8}Centre de Calcul de l'IN2P3, Villeurbanne, Lyon, France
\item \Idef{org9}Centro de Aplicaciones Tecnol\'{o}gicas y Desarrollo Nuclear (CEADEN), Havana, Cuba
\item \Idef{org10}Centro de Investigaci\'{o}n y de Estudios Avanzados (CINVESTAV), Mexico City and M\'{e}rida, Mexico
\item \Idef{org11}Centro Fermi - Museo Storico della Fisica e Centro Studi e Ricerche ``Enrico Fermi', Rome, Italy
\item \Idef{org12}Chicago State University, Chicago, Illinois, United States
\item \Idef{org13}China Institute of Atomic Energy, Beijing, China
\item \Idef{org14}Chonbuk National University, Jeonju, Republic of Korea
\item \Idef{org15}Comenius University Bratislava, Faculty of Mathematics, Physics and Informatics, Bratislava, Slovakia
\item \Idef{org16}COMSATS Institute of Information Technology (CIIT), Islamabad, Pakistan
\item \Idef{org17}Creighton University, Omaha, Nebraska, United States
\item \Idef{org18}Department of Physics, Aligarh Muslim University, Aligarh, India
\item \Idef{org19}Department of Physics, Ohio State University, Columbus, Ohio, United States
\item \Idef{org20}Department of Physics, Pusan National University, Pusan, Republic of Korea
\item \Idef{org21}Department of Physics, Sejong University, Seoul, Republic of Korea
\item \Idef{org22}Department of Physics, University of California, Berkeley, California, United States
\item \Idef{org23}Department of Physics, University of Oslo, Oslo, Norway
\item \Idef{org24}Department of Physics and Technology, University of Bergen, Bergen, Norway
\item \Idef{org25}Dipartimento di Fisica dell'Universit\`{a} 'La Sapienza' and Sezione INFN, Rome, Italy
\item \Idef{org26}Dipartimento di Fisica dell'Universit\`{a} and Sezione INFN, Cagliari, Italy
\item \Idef{org27}Dipartimento di Fisica dell'Universit\`{a} and Sezione INFN, Trieste, Italy
\item \Idef{org28}Dipartimento di Fisica dell'Universit\`{a} and Sezione INFN, Turin, Italy
\item \Idef{org29}Dipartimento di Fisica e Astronomia dell'Universit\`{a} and Sezione INFN, Bologna, Italy
\item \Idef{org30}Dipartimento di Fisica e Astronomia dell'Universit\`{a} and Sezione INFN, Catania, Italy
\item \Idef{org31}Dipartimento di Fisica e Astronomia dell'Universit\`{a} and Sezione INFN, Padova, Italy
\item \Idef{org32}Dipartimento di Fisica `E.R.~Caianiello' dell'Universit\`{a} and Gruppo Collegato INFN, Salerno, Italy
\item \Idef{org33}Dipartimento DISAT del Politecnico and Sezione INFN, Turin, Italy
\item \Idef{org34}Dipartimento di Scienze e Innovazione Tecnologica dell'Universit\`{a} del Piemonte Orientale and INFN Sezione di Torino, Alessandria, Italy
\item \Idef{org35}Dipartimento Interateneo di Fisica `M.~Merlin' and Sezione INFN, Bari, Italy
\item \Idef{org36}European Organization for Nuclear Research (CERN), Geneva, Switzerland
\item \Idef{org37}Faculty of Engineering and Business Administration, Western Norway University of Applied Sciences, Bergen, Norway
\item \Idef{org38}Faculty of Nuclear Sciences and Physical Engineering, Czech Technical University in Prague, Prague, Czech Republic
\item \Idef{org39}Faculty of Science, P.J.~\v{S}af\'{a}rik University, Ko\v{s}ice, Slovakia
\item \Idef{org40}Frankfurt Institute for Advanced Studies, Johann Wolfgang Goethe-Universit\"{a}t Frankfurt, Frankfurt, Germany
\item \Idef{org41}Gangneung-Wonju National University, Gangneung, Republic of Korea
\item \Idef{org42}Gauhati University, Department of Physics, Guwahati, India
\item \Idef{org43}Helmholtz-Institut f\"{u}r Strahlen- und Kernphysik, Rheinische Friedrich-Wilhelms-Universit\"{a}t Bonn, Bonn, Germany
\item \Idef{org44}Helsinki Institute of Physics (HIP), Helsinki, Finland
\item \Idef{org45}Hiroshima University, Hiroshima, Japan
\item \Idef{org46}Hochschule Worms, Zentrum  f\"{u}r Technologietransfer und Telekommunikation (ZTT), Worms, Germany
\item \Idef{org47}Horia Hulubei National Institute of Physics and Nuclear Engineering, Bucharest, Romania
\item \Idef{org48}Indian Institute of Technology Bombay (IIT), Mumbai, India
\item \Idef{org49}Indian Institute of Technology Indore, Indore, India
\item \Idef{org50}Indonesian Institute of Sciences, Jakarta, Indonesia
\item \Idef{org51}INFN, Laboratori Nazionali di Frascati, Frascati, Italy
\item \Idef{org52}INFN, Sezione di Bari, Bari, Italy
\item \Idef{org53}INFN, Sezione di Bologna, Bologna, Italy
\item \Idef{org54}INFN, Sezione di Cagliari, Cagliari, Italy
\item \Idef{org55}INFN, Sezione di Catania, Catania, Italy
\item \Idef{org56}INFN, Sezione di Padova, Padova, Italy
\item \Idef{org57}INFN, Sezione di Roma, Rome, Italy
\item \Idef{org58}INFN, Sezione di Torino, Turin, Italy
\item \Idef{org59}INFN, Sezione di Trieste, Trieste, Italy
\item \Idef{org60}Inha University, Incheon, Republic of Korea
\item \Idef{org61}Institut de Physique Nucl\'{e}aire d'Orsay (IPNO), Institut National de Physique Nucl\'{e}aire et de Physique des Particules (IN2P3/CNRS), Universit\'{e} de Paris-Sud, Universit\'{e} Paris-Saclay, Orsay, France
\item \Idef{org62}Institute for Nuclear Research, Academy of Sciences, Moscow, Russia
\item \Idef{org63}Institute for Subatomic Physics of Utrecht University, Utrecht, Netherlands
\item \Idef{org64}Institute for Theoretical and Experimental Physics, Moscow, Russia
\item \Idef{org65}Institute of Experimental Physics, Slovak Academy of Sciences, Ko\v{s}ice, Slovakia
\item \Idef{org66}Institute of Physics, Bhubaneswar, India
\item \Idef{org67}Institute of Physics of the Czech Academy of Sciences, Prague, Czech Republic
\item \Idef{org68}Institute of Space Science (ISS), Bucharest, Romania
\item \Idef{org69}Institut f\"{u}r Kernphysik, Johann Wolfgang Goethe-Universit\"{a}t Frankfurt, Frankfurt, Germany
\item \Idef{org70}Instituto de Ciencias Nucleares, Universidad Nacional Aut\'{o}noma de M\'{e}xico, Mexico City, Mexico
\item \Idef{org71}Instituto de F\'{i}sica, Universidade Federal do Rio Grande do Sul (UFRGS), Porto Alegre, Brazil
\item \Idef{org72}Instituto de F\'{\i}sica, Universidad Nacional Aut\'{o}noma de M\'{e}xico, Mexico City, Mexico
\item \Idef{org73}iThemba LABS, National Research Foundation, Somerset West, South Africa
\item \Idef{org74}Johann-Wolfgang-Goethe Universit\"{a}t Frankfurt Institut f\"{u}r Informatik, Fachbereich Informatik und Mathematik, Frankfurt, Germany
\item \Idef{org75}Joint Institute for Nuclear Research (JINR), Dubna, Russia
\item \Idef{org76}Konkuk University, Seoul, Republic of Korea
\item \Idef{org77}Korea Institute of Science and Technology Information, Daejeon, Republic of Korea
\item \Idef{org78}KTO Karatay University, Konya, Turkey
\item \Idef{org79}Laboratoire de Physique Subatomique et de Cosmologie, Universit\'{e} Grenoble-Alpes, CNRS-IN2P3, Grenoble, France
\item \Idef{org80}Lawrence Berkeley National Laboratory, Berkeley, California, United States
\item \Idef{org81}Lund University Department of Physics, Division of Particle Physics, Lund, Sweden
\item \Idef{org82}Nagasaki Institute of Applied Science, Nagasaki, Japan
\item \Idef{org83}Nara Women{'}s University (NWU), Nara, Japan
\item \Idef{org84}National and Kapodistrian University of Athens, School of Science, Department of Physics , Athens, Greece
\item \Idef{org85}National Centre for Nuclear Research, Warsaw, Poland
\item \Idef{org86}National Institute of Science Education and Research, HBNI, Jatni, India
\item \Idef{org87}National Nuclear Research Center, Baku, Azerbaijan
\item \Idef{org88}National Research Centre Kurchatov Institute, Moscow, Russia
\item \Idef{org89}Niels Bohr Institute, University of Copenhagen, Copenhagen, Denmark
\item \Idef{org90}Nikhef, National institute for subatomic physics, Amsterdam, Netherlands
\item \Idef{org91}NRC ¿Kurchatov Institute¿ ¿ IHEP , Protvino, Russia
\item \Idef{org92}NRNU Moscow Engineering Physics Institute, Moscow, Russia
\item \Idef{org93}Nuclear Physics Group, STFC Daresbury Laboratory, Daresbury, United Kingdom
\item \Idef{org94}Nuclear Physics Institute of the Czech Academy of Sciences, \v{R}e\v{z} u Prahy, Czech Republic
\item \Idef{org95}Oak Ridge National Laboratory, Oak Ridge, Tennessee, United States
\item \Idef{org96}Petersburg Nuclear Physics Institute, Gatchina, Russia
\item \Idef{org97}Physics department, Faculty of science, University of Zagreb, Zagreb, Croatia
\item \Idef{org98}Physics Department, Panjab University, Chandigarh, India
\item \Idef{org99}Physics Department, University of Jammu, Jammu, India
\item \Idef{org100}Physics Department, University of Rajasthan, Jaipur, India
\item \Idef{org101}Physikalisches Institut, Eberhard-Karls-Universit\"{a}t T\"{u}bingen, T\"{u}bingen, Germany
\item \Idef{org102}Physikalisches Institut, Ruprecht-Karls-Universit\"{a}t Heidelberg, Heidelberg, Germany
\item \Idef{org103}Physik Department, Technische Universit\"{a}t M\"{u}nchen, Munich, Germany
\item \Idef{org104}Research Division and ExtreMe Matter Institute EMMI, GSI Helmholtzzentrum f\"ur Schwerionenforschung GmbH, Darmstadt, Germany
\item \Idef{org105}Rudjer Bo\v{s}kovi\'{c} Institute, Zagreb, Croatia
\item \Idef{org106}Russian Federal Nuclear Center (VNIIEF), Sarov, Russia
\item \Idef{org107}Saha Institute of Nuclear Physics, Kolkata, India
\item \Idef{org108}School of Physics and Astronomy, University of Birmingham, Birmingham, United Kingdom
\item \Idef{org109}Secci\'{o}n F\'{\i}sica, Departamento de Ciencias, Pontificia Universidad Cat\'{o}lica del Per\'{u}, Lima, Peru
\item \Idef{org110}Stefan Meyer Institut f\"{u}r Subatomare Physik (SMI), Vienna, Austria
\item \Idef{org111}SUBATECH, IMT Atlantique, Universit\'{e} de Nantes, CNRS-IN2P3, Nantes, France
\item \Idef{org112}Suranaree University of Technology, Nakhon Ratchasima, Thailand
\item \Idef{org113}Technical University of Ko\v{s}ice, Ko\v{s}ice, Slovakia
\item \Idef{org114}Technische Universit\"{a}t M\"{u}nchen, Excellence Cluster 'Universe', Munich, Germany
\item \Idef{org115}The Henryk Niewodniczanski Institute of Nuclear Physics, Polish Academy of Sciences, Cracow, Poland
\item \Idef{org116}The University of Texas at Austin, Austin, Texas, United States
\item \Idef{org117}Universidad Aut\'{o}noma de Sinaloa, Culiac\'{a}n, Mexico
\item \Idef{org118}Universidade de S\~{a}o Paulo (USP), S\~{a}o Paulo, Brazil
\item \Idef{org119}Universidade Estadual de Campinas (UNICAMP), Campinas, Brazil
\item \Idef{org120}Universidade Federal do ABC, Santo Andre, Brazil
\item \Idef{org121}University College of Southeast Norway, Tonsberg, Norway
\item \Idef{org122}University of Cape Town, Cape Town, South Africa
\item \Idef{org123}University of Houston, Houston, Texas, United States
\item \Idef{org124}University of Jyv\"{a}skyl\"{a}, Jyv\"{a}skyl\"{a}, Finland
\item \Idef{org125}University of Liverpool, Liverpool, United Kingdom
\item \Idef{org126}University of Split, Faculty of Electrical Engineering, Mechanical Engineering and Naval Architecture, Split, Croatia
\item \Idef{org127}University of Tennessee, Knoxville, Tennessee, United States
\item \Idef{org128}University of the Witwatersrand, Johannesburg, South Africa
\item \Idef{org129}University of Tokyo, Tokyo, Japan
\item \Idef{org130}University of Tsukuba, Tsukuba, Japan
\item \Idef{org131}Universit\'{e} Clermont Auvergne, CNRS/IN2P3, LPC, Clermont-Ferrand, France
\item \Idef{org132}Universit\'{e} de Lyon, Universit\'{e} Lyon 1, CNRS/IN2P3, IPN-Lyon, Villeurbanne, Lyon, France
\item \Idef{org133}Universit\'{e} de Strasbourg, CNRS, IPHC UMR 7178, F-67000 Strasbourg, France, Strasbourg, France
\item \Idef{org134} Universit\'{e} Paris-Saclay Centre d¿\'Etudes de Saclay (CEA), IRFU, Department de Physique Nucl\'{e}aire (DPhN), Saclay, France
\item \Idef{org135}Universit\`{a} degli Studi di Pavia, Pavia, Italy
\item \Idef{org136}Universit\`{a} di Brescia, Brescia, Italy
\item \Idef{org137}V.~Fock Institute for Physics, St. Petersburg State University, St. Petersburg, Russia
\item \Idef{org138}Variable Energy Cyclotron Centre, Kolkata, India
\item \Idef{org139}Warsaw University of Technology, Warsaw, Poland
\item \Idef{org140}Wayne State University, Detroit, Michigan, United States
\item \Idef{org141}Westf\"{a}lische Wilhelms-Universit\"{a}t M\"{u}nster, Institut f\"{u}r Kernphysik, M\"{u}nster, Germany
\item \Idef{org142}Wigner Research Centre for Physics, Hungarian Academy of Sciences, Budapest, Hungary
\item \Idef{org143}Yale University, New Haven, Connecticut, United States
\item \Idef{org144}Yonsei University, Seoul, Republic of Korea
\end{Authlist}
\endgroup
  %%%%%%% done by webmaster team
\end{document}